\documentclass[preprint,amsmath,amssymb,
 aps, prd 
]{revtex4-1}

\usepackage{graphicx}
\usepackage{dcolumn}
\usepackage{bm}
\usepackage{soul}
\usepackage{hyperref}
\usepackage{color}

\begin{document}


\title{Anomalous and linear holographic hard wall models \\ for light unflavored mesons}

\author{Rafael A. Costa-Silva}
 \email{rafaelcosta@pos.if.ufrj.br}
 \email{rc-fis@outlook.com}
\author{Henrique Boschi-Filho}%
 \email{boschi@if.ufrj.br}
 \email{hboschi@gmail.com}
\affiliation{Instituto de Física, Universidade Federal do Rio de Janeiro, 21.941-909 - Rio de Janeiro - RJ - Brazil
}%

\date{\today}

\begin{abstract}
In this work we consider anomalous and linear holographic hard wall (HW) models for light unflavored mesons inspired by the AdS/CFT correspondence. The anomalous dimensions depend on the logarithm of the spin $S$ of the meson state and come from a semiclassical analysis of gauge/string duality. The anomalous HW model produces very good masses  and good Regge trajectories for mesons compared with PDG data. Inspired by this anomalous HW model we also propose a phenomenological modification of the dimension of the boundary operators such that the model produces asymptotic linear Regge trajectories. Both anomalous HW models considered here present mass percentage deviations of less than 3$\%$, when compared with PDG data. 
\end{abstract}

\maketitle


\section{Introduction}

The AdS/CFT correspondence \cite{Aharony:1999ti} establishes  relations between Yang-Mills (YM) and higher dimensional gravitational theories. One of the amazing properties of this  correspondence is that if the YM theories are strongly coupled, the dual gravitational one is weak coupled allowing for the solution of a myriad of phenomena from the quark-gluon plasma to condensed matter (for reviews, see {\sl e.g.}, \cite{Casalderrey-Solana:2011dxg, Zaanen:2015oix}). In particular, hadronic spectra were described by different proposals such as the holographic hard wall (HW) and  the soft wall (SW) models \cite{Polchinski:2001tt, Polchinski:2002jw, Boschi-Filho:2002xih, Boschi-Filho:2002wdj, deTeramond:2005su, Erlich:2005qh, Boschi-Filho:2005xct, Karch:2006pv, Colangelo:2008us}. The HW model \cite{Boschi-Filho:2002xih, Boschi-Filho:2002wdj, deTeramond:2005su, Erlich:2005qh, Boschi-Filho:2005xct} is well known for producing reasonable hadronic masses but non-linear (parabolic) Regge trajectories. On the other side, the SW model \cite{Karch:2006pv, Colangelo:2008us} also give good masses with the advantage of generating linear Regge trajectories. For other interesting QCD properties described by the HW model, see {\sl e.g.} \cite{Boschi-Filho:2005nmp, Boschi-Filho:2006hfm, Herzog:2006ra, BallonBayona:2007vp, Grigoryan:2007my, Kwee:2007dd, Grigoryan:2007wn, Jo:2009xr, Costa:2013uia, Craps:2013iaa, Mamo:2015dea, Ballon-Bayona:2017bwk}. 

In this work, we are going to consider the inclusion of anomalous dimensions for the boundary operators describing light unflavored mesons with spins $S\ge 1$ in the HW model. These anomalous dimensions depend on the logarithm of the spin $S$ of the mesonic state inspired by a semiclasical analysis of gauge/string dualities \cite{Gubser:2002tv} and was used recently to discuss the case of even spin glueballs \cite{Costa-Silva:2023vuu}. We find very good masses and Regge trajectories for the studied mesons when compared with PDG data \cite{PDG:2022pth}, better than the ones from the HW and SW models. Furthermore, we discuss a phenomenological modification of the dimension of the boundary operators representing the mesonic states in order to obtain asymptotically linear Regge trajectories, similar to the case of glueballs. Actually, the linearization of the mesonic states require an additional parameter to obtain the best fit to PDG data. For a recent account on non-holographic meson Regge trajectories, see {\sl e.g.} \cite{Chen:2021kfw}. For later reference, we present in Table \ref{tabpdg} the masses of the meson families starting with the states  $\rho(770)$, $\omega(782)$, $b_1(1235)$, and $h_1(1170)$, with their corresponding Regge trajectories. These families are classified according to their $PC$ parity (even when $P$ and $C$ have the same sign or odd with opposite signs) and isospins $I=0,1$. Note that, according to PDG notation, mesons are classified as $J^{PC}$, where $J$ is the total angular momenta, $P$ is the parity and $C$ the charge conjugation of the state. In our notation, we simply call the total angular momenta as the ``spin'' $S$, so that $J=S$. The $G$ parity ($G=(-)^{I+S}$) of these mesons is also included in this table following PDG notation but is not relevant in the present analysis.

\begin{table}[ht!]
\vskip 0.5cm
\centering
\begin{tabular}{|c|c|c|c|c|c|}
\hline
\hline
$I^G(J^{PC})$ &  Meson & PDG (MeV) &  PDG Regge trajetories (GeV)\\ \hline
 $1^{+}(1^{--})$   & $\rho (770)$ & $775.26 \pm 0.25$ & \\
 
 $1^{-}(2^{++})$  & $a_2 (1320)$&$1316.9\pm 0.9$ & \\
 
 $1^{+}(3^{--})$ & $\rho_3 (1690)$&$1688.8\pm2.1$ & $J=(0.88\pm0.04)M^2 +0.47\pm0.15$\\
 
 $1^{-}(4^{++})$ & $a_4 (1970)$&$1967\pm16$ & \\
 
 $1^{+}(5^{--})$ & $\rho_5 (2350)$ &$2330\pm35$ & \\
 
 $1^{-}(6^{++})$ & $a_6 (2450)$&$2450\pm130$ & \\
\hline
 $0^{-}(1^{--})$ & $\omega (782)$ & $782.66\pm0.13$ & \\
 
 $0^{+}(2^{++})$& $f_2 (1270)$&$1275.5\pm 0.8$ & \\
 
 $0^{-}(3^{--})$& $\omega_3 (1670)$&$1667\pm4$ & $J=(0.90\pm0.02)M^2 +0.47\pm0.07$\\
 
 $0^{+}(4^{++})$& $f_4 (2050)$&$2018\pm11$ & \\
 
 $0^{-}(5^{--})$& $\omega_5 (2250)$ & $2250\pm70$ & \\
 
 $0^{+}(6^{++})$ & $f_6 (2510)$&$2465\pm50$ & \\ 
 \hline
 $1^{+}(1^{+-})$& $b_1 (1235)$&$1229.5\pm 3.2$ & \\
 
 $1^{-}(2^{-+})$ & $\pi_2 (1670)$&$1670.6\pm2.05$ & $J=(0.83\pm0.04)M^2-0.30\pm0.16$\\
 
 $1^{+}(3^{+-})$& $b_3 (2030)$&$2032\pm12$ & \\
 
 $1^{-}(4^{-+})$ & $\pi_4 (2250)$ & $2250\pm15$ & \\ 
\hline
 $0^{-}(1^{+-})$ & $h_1 (1170)$&$1166\pm6$ & \\
 
 $0^{+}(2^{-+})$ & $\eta_2 (1645)$&$1617\pm5$ & $ J=(0.73\pm0.02)M^2+0.03\pm0.06$\\
 
 $0^{-}(3^{+-})$& $h_3 (2025)$&$2025\pm20$ & \\
 
 $0^{+}(4^{-+})$& $\eta_4 (2330)$ & $2328\pm38$ & \\
 \hline\hline
\end{tabular}
\vskip 0.5cm
\caption{\label{tabpdg} Masses in MeV for light unflavored mesons  classified under even or odd $PC$, with isospins $I=0,1$, and spins $J=S\ge1$  from PDG \cite{PDG:2022pth}, and their corresponding Regge trajectories in GeV.}
\end{table}

This work is organized as follows. In Section \ref{OHW}, we review the HW model for light unflavored mesons and in Section \ref{AHW} we discuss the introduction of anomalous dimensions in the HW model and obtain the spectra for the mesonic families starting with the states of the $\rho(770)$, $\omega(782)$, $b_1(1235)$, and $h_1(1170)$. Then, in Section \ref{LHW} we modify the dimension of the mesonic boundary operators such that we find asymptotically linear Regge trajectories for the mesons already discussed in the previous section. Finally, in Section \ref{conclusions} we present our conclusions.

\section{Original HW model for mesons}\label{OHW}

In this work we are interested in studying the spectra of light unflavored mesons, and their Regge trajectories using anomalous and linear holographic HW models. These mesons can be classified in four families according to their behavior under ${PC}$ transformations (even or odd), isospin $I=0$ or $1$, and spins $S\ge 1$. These four families imply four distinct Regge trajectories.  

In order to describe such objects using the holographic HW model, we need to study a massive vector field in $AdS_5$. We start with the action
\cite{Aharony:1999ti}
\begin{equation}\label{action}
    S=-\frac{1}{2}\int d^5x \sqrt{-g}\bigg[\frac{1}{2}g^{pm}g^{qn}F_{mn}F_{pq}+M_5^2g^{pm}A_pA_m\bigg],
\end{equation}
where $F_{mn}=\partial_mA_n-\partial_nA_m$ is the stress tensor of the Proca field $A_m$ with mass $M_5$ in the five-dimensional AdS space. Moreover, $g$ is the determinant of the metric of $AdS_5$,
\begin{equation}\label{adsmetric}
    ds^2=g_{mn}dx^mdx^n=\frac{R^2}{z^2}(dz^2+\eta_{\mu\nu}dx^{\mu}dx^{\nu}), 
\end{equation}
with $R$ being the $AdS$ radius, $z$ the holographic coordinate and $\eta_{\mu\nu}$ is the Minkowski four-dimensional boundary metric. The equation of motion that follows from the action, Eq. (\ref{action}), is given by 
\begin{equation}\label{eom1}
    \partial_z\bigg[\bigg(\frac{1}{z}\bigg)F_{zn}\eta^{nq}\bigg]+\partial_{\mu}\bigg[\bigg(\frac{1}{z}\bigg)\eta^{m\mu}F_{mn}\eta^{nq}\bigg]-\frac{(M_5R)^2}{z^3}A_n\eta^{nq}=0.
\end{equation}
We use a plane wave solution as an ansatz for the five-dimensional vector field $A_m(z,x^\mu)$ with four-momentum $q_\mu$, and polarization four-vector $\epsilon_{\rho}$,  
\begin{eqnarray}\label{ansatz}
    &&A_{\rho}(z,x^{\mu})=\epsilon_{\rho}v(z)e^{iq_{\mu}x^{\mu}}, \\
    &&A_z=0, 
\end{eqnarray}
with $\partial_{\mu}A^{\mu}=0$. Defining  
 $v(z)=z f(z)$, one obtains 
\begin{equation}\label{eom2}
    z^2\frac{d^2\psi(z)}{dz^2}+z\frac{d\psi(z)}{dz}-q^2z^2\psi(z)-[1+(M_5R)^2] f(z)=0.
\end{equation}
This is Bessel equation, whose solutions are linear combinations of Bessel ($J_{\nu}$) and Neumann ($Y_{\nu})$ functions both of order $\nu=\sqrt{1+(M_5R)^2}$. Since we are interested only in regular solutions in the bulk, we will consider just the Bessel solution
\begin{equation}\label{f(z)solution}
    f(z)\sim  J_{\nu}(m z),
\end{equation}
where $m$ is an irrelevant constant in the conformal field theory defined on the Minkowski boundary of the $AdS_5$ space, due to scale invariance, but has a fundamental role for the HW model with the introduction of the cutoff in the AdS space (see Eq. \eqref{zmax} below).

On the other hand, the mesons with spin $S=1,2,3,...$, that we are interested in are described by operators in the boundary Minkowski space as \cite{Gubser:2002tv} 
 \begin{equation}\label{mesonOp}
     {\cal O}_{\mu_1...\mu_S}^{3+S}=
     \;\bar\psi 
     D_{\{\mu_1\dots}D_{\mu_S\}}\psi\;;
     \qquad(S=1, 2, 3, ...)\,,
 \end{equation}
 where $\psi$ and $\bar \psi$ represent quark and antiquark fields while $D_{\mu_i}$ with $i=1, \dots, S$, are symmetrized covariant derivatives \cite{Operador}. Then, the canonical dimensions of these operators are given by
\begin{equation}\label{canonicaldim}
     \Delta=\Delta_{\rm can.}=3+S. 
 \end{equation}

The link between the operators $\cal O$  and string excitations in  $AdS_5$ can be constructed from its conformal dimension. From the AdS/CFT correspondence, the vector field in $AdS_5$ that we are describing is a $1$-form, so its holographic mass $M_5$ is related to the conformal dimension $\Delta$ of boundary operators $\cal O$, and given by \cite{Aharony:1999ti}
\begin{equation}\label{holographicmass}
    (M_5R)^2=(\Delta-1)(\Delta-3).
\end{equation}
Using this relation in the index $\nu=\sqrt{1+(M_5R)^2}$ of the Bessel function, one sees that
\begin{equation}\label{nuDelta}
    \nu=\Delta-2.
\end{equation}
 Furthermore, combining Eqs. (\ref{nuDelta}) and (\ref{canonicaldim}), one finds that the Bessel functions describing the mesons represented  by the operators $\cal O$, Eq. \eqref{mesonOp}, have order
\begin{equation}\label{nuS}
    \nu=S+1.
\end{equation}

    The HW model consists in the holographic relation between bulk fields in AdS space with an infrared (IR) cutoff that breaks the conformal symmetry of the boundary theory. That means a hard cutoff in the holographic $z$ coordinate, so we are working in an AdS slice, given by $0\leq z\leq z_{\rm max}$, where
    \begin{equation}\label{zmax}
        z_{\rm max}=\frac{1}{\Lambda_{\rm HW}}, 
    \end{equation}
with $\Lambda_{\rm HW}$ a convenient energy scale in the HW model to be fixed later (see Eq. \eqref{massgdst} below). 
Then, the solution of the vector bulk field, Eq. \eqref{f(z)solution}, assumes the form
\begin{equation}\label{psisolution}
    f(z)\sim J_{\nu}(m_{\nu,k}\,z),
\end{equation}
where $m_{\nu,k}$ is identified with the mass of the boundary operator $\cal O$ that lives in the four-dimensional Minkowski space, describing  mesons. Note that $k=1,2,3...,$ denote radial excitations, where $k=1$ is the corresponding ground state. In this work we are interested in angular momenta excitations, so we take $k=1$ for all the objects from now on.

\begin{table}[ht!]
\centering
\begin{tabular}{|c|c|c|c|c|c|c|c|}
\hline
\hline
$I^{G}(J^{PC})$& Meson & HW  &$\delta_{\rm HW}$& $\Lambda_{\rm HW}$ & HW Regge Trajectories (GeV)& SW  &$\delta_{\rm SW}$
\\ \hline
 $1^{+}(1^{--})$ &$\rho (770)$ &$775.26$&$0\%$ & & &$775.26$&$0\%$
 \\
 $1^{-}(2^{++})$&$a_2 (1320)$&$963.13$ &$25.9\%$& & & $1096.38$&$16.7\%$
 \\
 $1^{+}(3^{--})$&$\rho_3 (1690)$&$1145.52$&$32.2\%$&150.96 & 
$J=(2.25\pm0.10)M^2-0.12\pm0.19$ &
 $1342.79$&$20.5\%$
 \\
 $1^{-}(4^{++})$&$a_4 (1970)$&$1324.12$&$32.7\%$&&  & 
 $1550.52$&$21.2\%$
 \\
 $1^{+}(5^{--})$& $\rho_5 (2350)$ & $1499.93$&$35.6\%$&
 &
 &
 $1733.53$&$25.6\%$
 \\
 $1^{-}(6^{++})$& $a_6 (2450)$&$1673.57$&$31.7\%$&&
 &
 $1898.99$&$22.5\%$
\\
\hline
 $0^{-}(1^{--})$&$\omega (782)$ & $782.66$&$0\%$&
 &
 &
 $782.66$&$0\%$
 \\
 $0^{+}(2^{++})$&$f_2 (1270)$&$972.31$&$23.8\%$&
 &
 &
 $1106.85$&$13.2\%$
 \\
 $0^{-}(3^{--})$&$\omega_3 (1670)$&$1156.44$&$30.6\%$
 &
152.24
&
$J=(2.21\pm0.10)M^2-0.12\pm0.19$  &$1355.61$&$18.7\%$
 \\
 $0^{+}(4^{++})$&$f_4 (2050)$&$1336.74$&$33.8\%$
 &
 &
 &$1565.32$&$22.4\%$
 \\
 $0^{-}(5^{--})$&$\omega_5 (2250)$ & $1514.23$&$32.7\%$&
 &
 &
 $1750.08$&$22.2\%$
 \\
 $0^{+}(6^{++})$&$f_6 (2510)$&$1689.52$&$31.5\%$
 &
 &
 &$1917.12$&$22.2\%$
\\ \hline
 $1^{+}(1^{+-})$&$b_1 (1235)$&$1229.5$&$0\%$
 &
 &
 &$1229.5$&$0\%$
 \\
 $1^{-}(2^{-+})$&$\pi_2 (1670)$&$1527.41$&$8.57\%$
 &
239.41 
& $J=(1.03\pm0.05)M^2-0.48\pm0.15$
 & $1738.78$&$4.08\%$
 \\
 $1^{+}(3^{+-})$& $b_3 (2030)$&$1816.70$&$10.6\%$
 &
 &
 &$2129.56$&$4.80\%$
 \\
 $1^{-}(4^{-+})$& $\pi_4 (2250)$ & $2099.95$&$6.67\%$
 &
 &
 &$2459.00$&$9.29\%$
\\ \hline
 $0^{-}(1^{+-})$&$h_1 (1170)$&$1166.0$&$0\%$
 &
 &
 &$1166$&$0\%$
 \\
 $0^{+}(2^{-+})$&$\eta_2 (1645)$&$1448.56$&$10.4\%$
 &
227.04
& $J=(1.15\pm0.05)M^2-0.48\pm0.15$ 
 &$1648.97$&$1.98\%$
 \\
 $0^{-}(3^{+-})$&$h_3 (2025)$&$1722.87$&$14.9\%$
 &
 &
 &$2019.57$&$0.27\%$
 \\
 $0^{+}(4^{-+})$&$\eta_4 (2330)$ &$1991.49$&$14.5\%$
 &
 &
 &$2332.00$&$0.17\%$\\
 \hline\hline
\end{tabular}
\vskip 0.5cm
\caption{\label{HWSW} Masses in MeV for light unflavored mesons  classified under even or odd $PC$, with isospins $I=0,1$, and spins $J=S\ge1$  from the HW model, Eq. \eqref{masstower},  and their corresponding Regge trajectories in GeV, and $\Lambda_{\rm HW}$ in MeV for each family. For comparison, we also show masses from the SW model, according to $m_{S}^2\propto S$ \cite{Karch:2006pv}. The percentage deviations of both models refer to PDG data presented in Table \ref{tabpdg}.}
\end{table}

 Imposing Dirichlet boundary conditions (for $k=1$) in Eq. (\ref{psisolution}) at $z=z_{\rm max}$, one obtains 
    \begin{equation}
        J_{\nu}\bigg(\frac{m_{\nu,1}}{\Lambda_{\rm HW}}\bigg)=0.
    \end{equation}
    Defining the first zero $\chi_{\nu,1}$ of the Bessel function of order $\nu$, such that  $J_{\nu}(\chi_{\nu,1})=0$, and making use of the relation Eq. (\ref{nuS}), we can write the mass of the mesonic state with spin $S$ as 
\begin{equation}\label{masstower}
   m_{\nu,1}=\chi_{\nu,1}\Lambda_{\rm HW}; \qquad\qquad (\nu=S+1).
    \end{equation}
In particular, from the state with $S=1$ we fix
\begin{equation}\label{massgdst}
        \Lambda_{\rm HW}= \frac{m_{2,1}}{\chi_{2,1}}, 
    \end{equation}
 taking $m_{2,1}$ as an input from PDG \cite{PDG:2022pth} for the states $\rho (770)$, $\omega (782)$, $b_1 (1235)$, and $h_1 (1170)$. 
The meson masses obtained from the HW model, using Eq. \eqref{masstower}, are shown in Table \ref{HWSW}, with the  corresponding values of $\Lambda_{\rm HW}$ and Regge trajectories found for each mesonic family, together with the SW model predictions with the relative deviations with respect to PDG data, presented in Table \ref{tabpdg}.

\begin{figure}[h]
\vskip0.5cm 
\centering
\includegraphics[width=7.5cm]{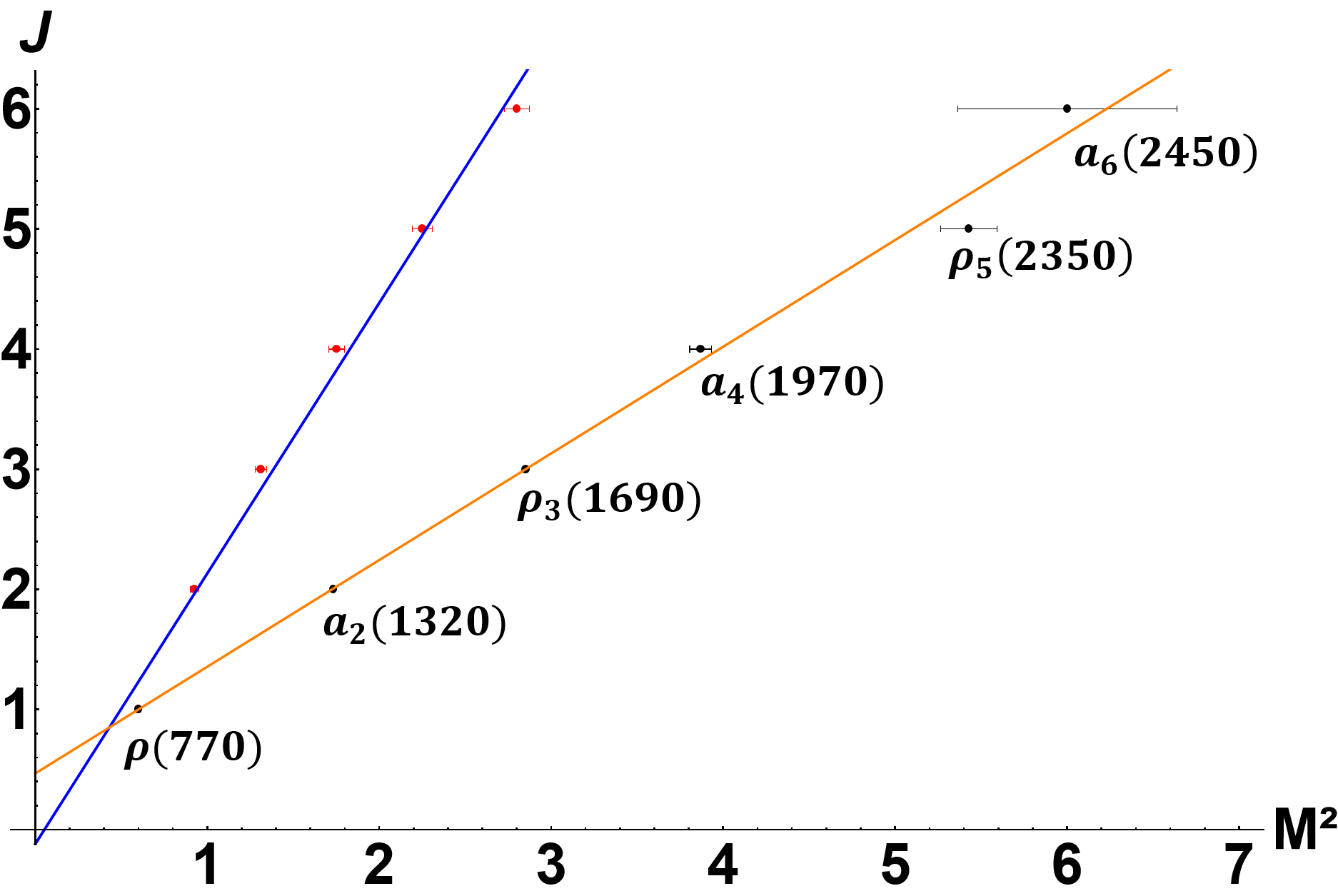}\qquad 
\includegraphics[width=7.5cm]{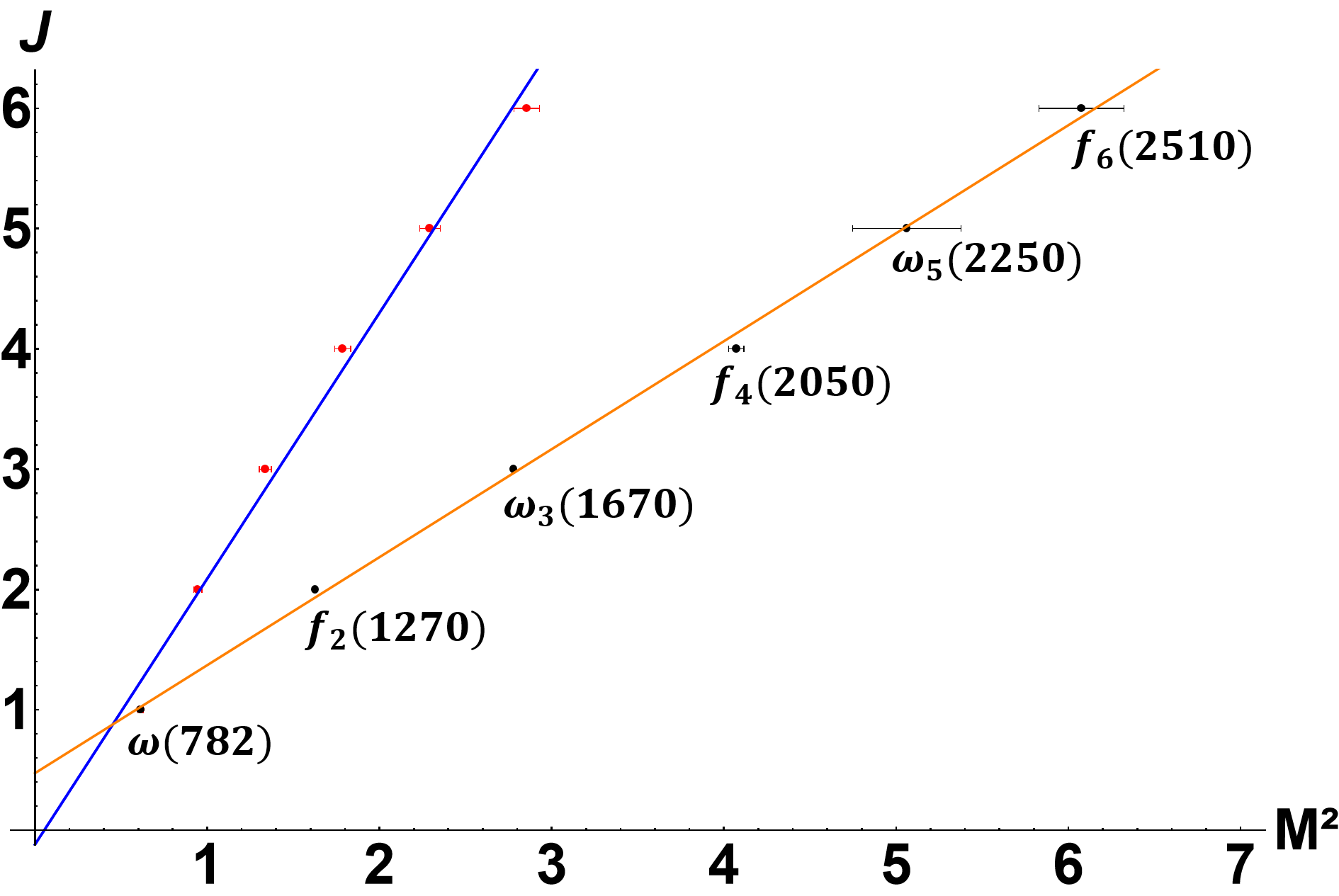}\vskip 0.5cm 
\includegraphics[width=7.5cm]{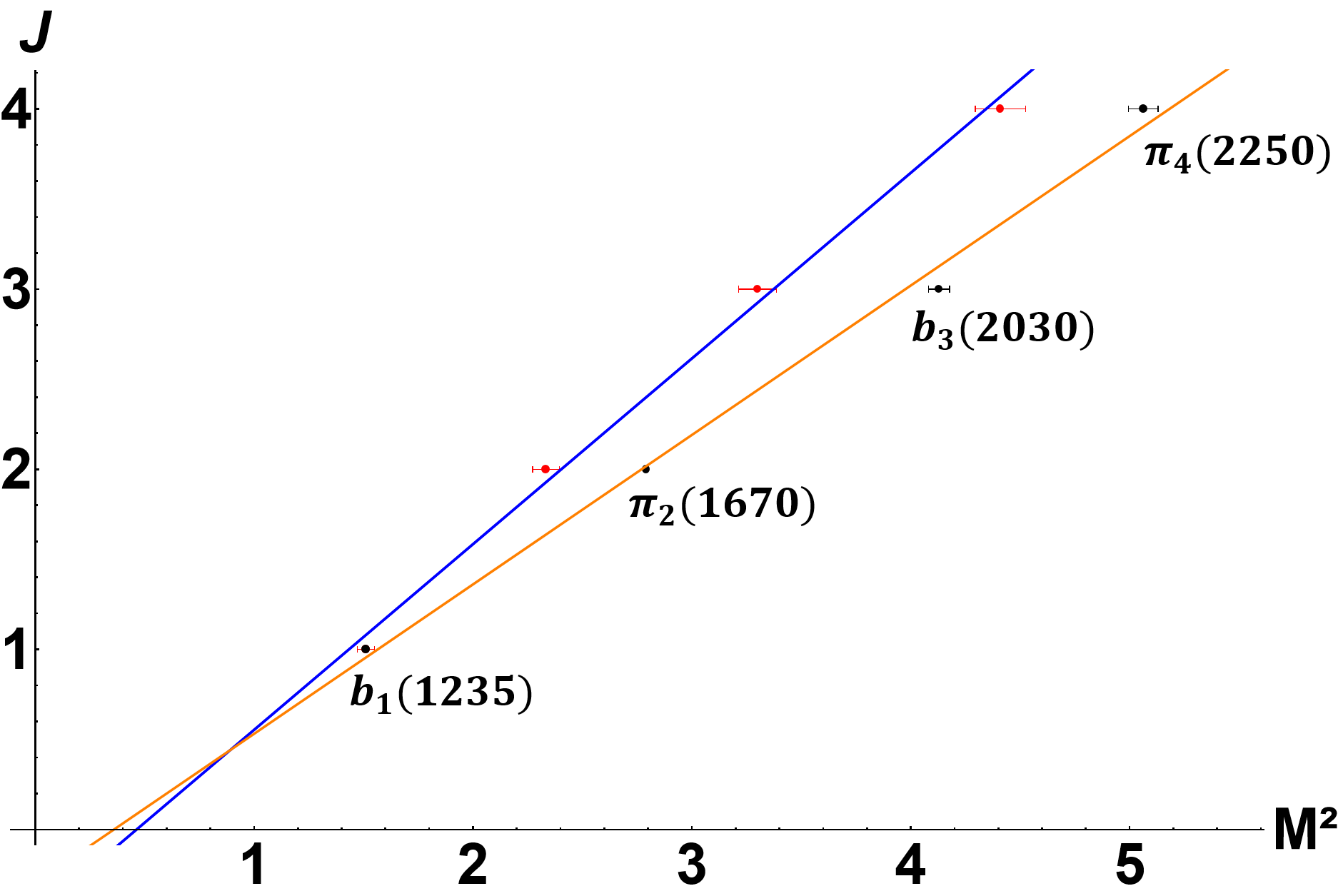}\qquad 
\includegraphics[width=7.5cm]{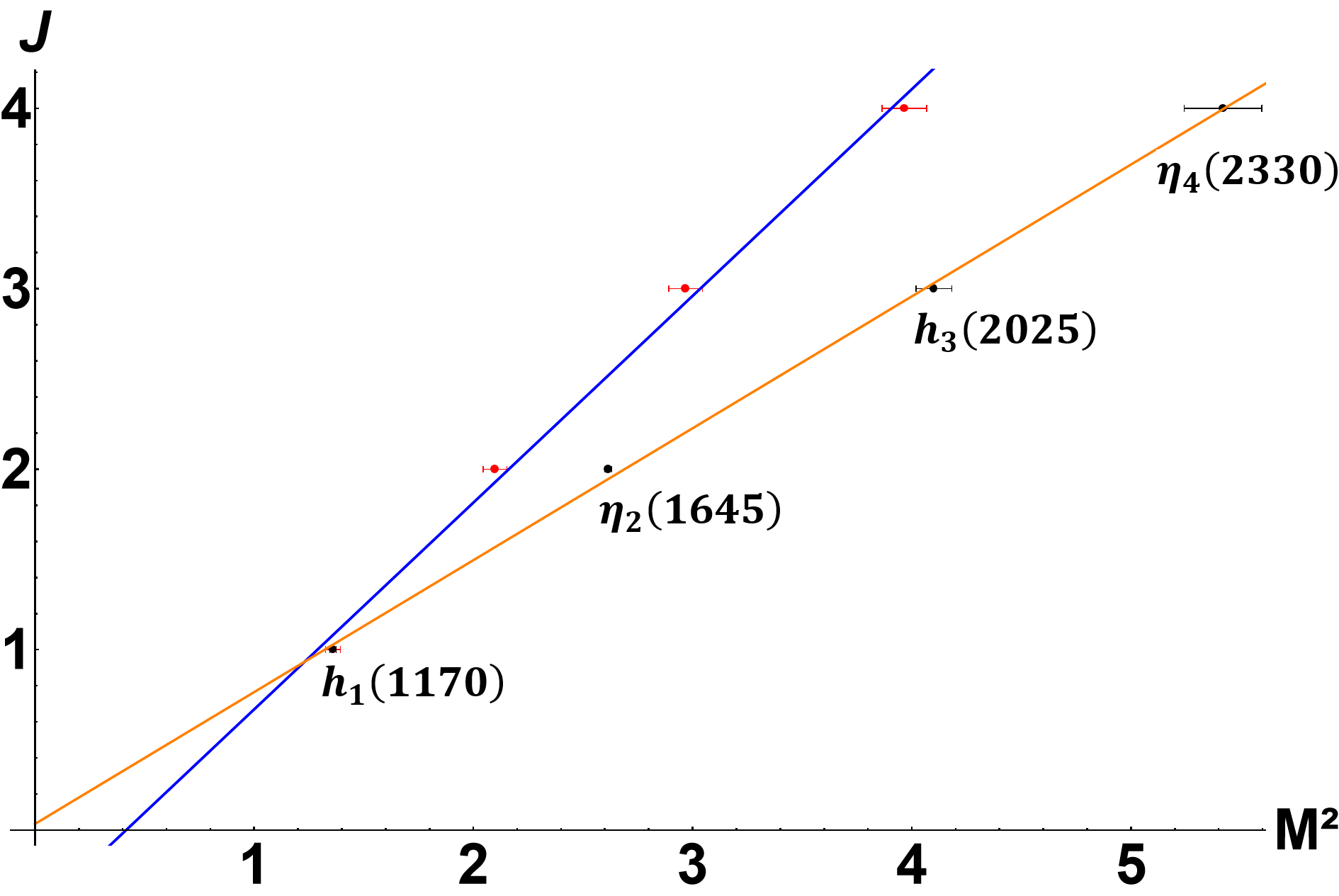}
\caption{Regge trajectories ($J\times M^2$)  from the HW  model (blue lines) and from PDG data (orange lines) 
for mesons, with masses expressed in GeV. Red dots are the values obtained from Eq. \eqref{masstower} in the HW  model with uncertainties shown in Table \ref{HWSW}, while the black dots correspond to PDG data shown in Table \ref{tabpdg}. The blue and orange lines are  obtained by linear regression of these points. {\sl Upper left panel:} mesons with even ${PC}$, $I=1$ ($\rho (770)$ family); {\sl Upper right panel:}  mesons with even ${PC}$, $I=0$ ($\omega (782)$ family); {\sl Lower left panel:}  mesons with odd ${PC}$, $I=1$ ($b_1 (1235)$ family); {\sl Lower right panel:}  mesons with odd ${PC}$, $I=0$ ($h_1 (1170)$ family).}
\label{figHW}
\end{figure}

    The process described above is known as the holographic HW  model and was used in the past to obtain hadronic  spectra among other properties \cite{Polchinski:2001tt, Polchinski:2002jw, Boschi-Filho:2002xih, Boschi-Filho:2002wdj, deTeramond:2005su, Erlich:2005qh, Boschi-Filho:2005xct, Boschi-Filho:2005nmp, Boschi-Filho:2006hfm, Herzog:2006ra, BallonBayona:2007vp, Grigoryan:2007my, Kwee:2007dd, Grigoryan:2007wn, Jo:2009xr, Costa:2013uia, Craps:2013iaa, Mamo:2015dea, Ballon-Bayona:2017bwk}. Nevertheless,  it is a well-known fact that this model implies non-linear Regge trajectories \cite{Boschi-Filho:2002xih, Boschi-Filho:2002wdj, deTeramond:2005su, Erlich:2005qh, Boschi-Filho:2005xct, Karch:2006pv} in contrast with experimental data 
(see, {\sl e.g.},  \cite{Spector:1967xsn, Kanki:1975fp})

For comparison, we plot in Figure \ref{figHW}  experimental data (from Table \ref{tabpdg}) for the meson families starting with the states  $\rho (770)$, $\omega (782)$, $b_1 (1235)$, and $h_1 (1170)$ together with HW masses (shown in Table \ref{HWSW}) with uncertainties and the corresponding Regge trajectories obtained by linear regression of these points. The equations for these trajectories are also shown in Tables 
\ref{tabpdg} and \ref{HWSW}. As one can see from this Figure, the agreement between HW and PDG masses is not very good, specially for the $\rho(770)$ and $\omega(782)$ families.

    In the next section we develop an improvement of the HW model, by considering anomalous dimensions, in order to obtain better masses compared with the PDG ones, and  Regge trajectories for these meson families which are closer to experimental data.





\section{Anomalous HW model}\label{AHW}

  In an interacting theory, the dimension of an operator is no longer canonical, due to renormalization that introduces   anomalous contributions  ($\Delta_{\rm Anom}$). So, the total dimension of an operator is given by its canonical plus the anomalous dimensions: 
\begin{equation}\label{totaldelta}
        \Delta_{\rm Total}=\Delta_{\rm can.}+\Delta_{\rm Anom}\,.
    \end{equation} 
    
    From a well known semi-classical gauge/string duality analysis \cite{Gubser:2002tv}, it is found that the anomalous dimension increases logarithmically with the spin $S$ of the boundary operators as
    \begin{equation}
        \Delta_{\rm Anom}=\frac{\sqrt{\lambda}}{\pi}\ln{\bigg(\frac{S}{\sqrt{\lambda}}\bigg)}+\mathcal{O}(S^0),
    \end{equation}
    where $\lambda$ is the 't Hooft coupling. Inspired by this result, we consider an anomalous dimension given by
        \begin{equation}\label{anomdim}
             \Delta_{\rm Log}= a_{\rm Log}\ln(S),
        \end{equation}
    where $a_{\rm Log}$ is a constant that will be determined by best fit for each meson family. 
    
     Then, the anomalous HW (AHW) model consists in considering bulk fields in AdS slice, Eq. \eqref{zmax}, with  total dimension given by Eq. \eqref{totaldelta} such that 
      the mass associated with boundary operators, now takes the form
\begin{equation}\label{masstower2}
        m_{\nu,1} = \chi_{\nu,1} \Lambda_{\rm HW}; \qquad\qquad (\nu=S+1+\Delta_{\rm Log}), 
    \end{equation}
where we used the fact that $\Delta_{\rm can.}=3+S$ for these mesonic states and that the order of the corresponding Bessel function is $\nu=\Delta_{\rm Total}-2$. The value of $\Lambda_{\rm HW}$ is determined for each meson familiy using as inputs the masses of the states with $S=1$ from PDG, namely $\rho (770)$, $\omega (782)$, $b_1 (1235)$, and $h_1 (1170)$. By construction, these values of $\Lambda_{\rm HW}$ are the same as those ones used in the usual HW model, listed in Table \ref{HWSW}, since Eq. \eqref{anomdim} implies a vanishing anomalous dimension $\Delta_{\rm Log}$ for $S=1$. 
    
    In order to obtain the best numerical fit, let us define the fractional deviation:
    \begin{equation}\label{percentualdev}
        \delta_i=\frac{|m_i-M_i|}{M_i},
    \end{equation}
where $m_i$ is the mass of a determined particle obtained by our model, from Eq. \eqref{masstower2}, and $M_i$ its mass obtained from PDG, listed in Table \ref{tabpdg}. Further, we define the function $Q$ as
    \begin{equation}\label{minimization}
        Q=\sqrt{\sum_i \delta_i^2}.
    \end{equation}
So, the best fit method adopted here corresponds to minimization of the deviations from PDG data given by the function $Q$ for each meson family in the AHW model.

\begin{table*}[t!]
\begin{tabular}{|c|c|c|c|c|c|c|}
\hline 
\hline 
$I^{G}(J^{PC})$ & Meson & AHW Log &$\delta_{\rm Log}$&$\Delta_{\rm Log}$ 
\\ \hline
 $1^{+}(1^{--})$ & $\rho (770)$&$775.26\pm10.08$&$0\%$&$0$ 
 \\
 $1^{-}(2^{++})$ & $a_2 (1320)$&$1309.95\pm22$&$0.53\%$&$1.92$
 \\
 $1^{+}(3^{--})$& $\rho_3 (1690)$&$1681.02\pm31$&$0.46\%$&$3.04$
 \\
 $1^{-}(4^{++})$ & $a_4 (1970)$&$1988.75\pm37$&$1.11\%$&$3.84$
 \\
 $1^{+}(5^{--})$& $\rho_5 (2350)$&$2262.37\pm43$&$2.90\%$&$4.46$
 \\
 $1^{-}(6^{++})$& $a_6 (2450)$&$2514.58\pm47$&$2.64\%$&$4.96$
\\ \hline
 $0^{-}(1^{--})$ & $\omega (782)$&$782.66\pm10.17$&$0\%$&$0$
 \\
 $0^{+}(2^{++})$&$f_2 (1270)$&$1302.58\pm22$&$2.12\%$&$1.81$
 \\
 $0^{-}(3^{--})$&$\omega_3 (1670)$&$1666.4\pm30$&$0.04\%$&$2.87$
 \\
 $0^{+}(4^{++})$& $f_4 (2050)$&$1969.63\pm37$&$2.40\%$&$3.62$
 \\
 $0^{-}(5^{--})$& $\omega_5 (2250)$&$2240.19\pm42$&$0.44\%$&$4.20$
 \\
 $0^{+}(6^{++})$& $f_6 (2510)$&$2490.23\pm46$&$1.02\%$&$4.68$
\\ \hline
 $1^{+}(1^{+-})$& $b_1 (1235)$&$1229.5\pm16$&$0\%$&$0$
\\
 $1^{-}(2^{-+})$&$\pi_2 (1670)$&$1640.57\pm27$&$1.80\%$&$0.39$
 \\
 $1^{+}(3^{+-})$ & $b_3 (2030)$&$1991.55\pm37$&$1.99\%$&$0.62$
 \\
 $1^{-}(4^{-+})$& $\pi_4 (2250)$&$2316.73\pm44$&$2.97\%$&$0.78$
\\ \hline
 $0^{-}(1^{+-})$ & $h_1 (1170)$&$1166\pm15$&$0\%$&$0$
 \\
 $0^{+}(2^{-+})$ & $\eta_2 (1645)$&$1629.93\pm25$ &$0.80\%$ & $0.66$  
 \\
 $0^{-}(3^{+-})$ & $h_3 (2025)$&$2003.12\pm33$&$1.08\%$&$1.04$
 \\
 $0^{+}(4^{-+})$& $\eta_4 (2330)$&$2339.01\pm39$&$0.47\%$&$1.32$ 
 \\
 \hline 
 \hline 
\end{tabular}
\caption{\label{HWlog} Masses in MeV for mesons  classified under even or odd $PC$  with isospins  $I=0,1$ and spins $S\ge1$, from the AHW Log model, determined from Eq. \eqref{masstower2}, and the percentage deviations $\delta_{\rm Log}=\delta_{i}\times 100\%$, with $\delta_{i}$ given by Eq. \eqref{percentualdev}, of the masses from this model compared with PDG data, using  minimization with respect to the  parameter $Q$, Eq. \eqref{minimization}. For each family the mass of the state with $J=S=1$  is taken as an input from PDG. In the last column we show the anomalous dimension $\Delta_{\rm Log}$ for each state determined by Eq. \eqref{anomdim}.}
\end{table*}

Further, we want to know by how much these masses are close to the linear Regge trajectory, obtained from the model, for each family. In order to measure this, we compute 
\begin{equation}\label{Rquadrado}
    R^2=1-\frac{\sum_i [m_i-\mathcal{M}_i(S)]^2}{\sum_i (m_i - \bar{m})^2},
\end{equation}
where $m_i$ is the mass obtained from the  AHW model, Eq. \eqref{masstower2}, for each state $i$, while ${\cal M}_i(S)$ is the corresponding mass of this state with angular momentum $S$, computed from the equation of the Regge trajectory and $\bar{m}$ is the average of the masses $m_i$. In a perfect model one would have  $R^2=1$, so the closer $R^2$ is to one, the best is the  match between the output masses of the model and the linear Regge trajectory.

     We also consider an uncertainty for the parameter $a_{\rm Log}$, defined in Eq. \eqref{anomdim}, which implies uncertainties for the masses ${m_i}$, obtained by error propagation, according to the formula:
    \begin{eqnarray}\label{uncert}
        \delta{m_i}=\sqrt{\left(\frac{\chi_{\nu_i,1}}{\chi_{2,1}}\right)^2 \left(\delta m_0\right)^2 + \left(\frac{m_0}{\chi_{2,1}}\right)^2 
        \left(\delta \chi_{\nu_i,1}\right)^2}, 
    \end{eqnarray}
where $\delta m_0$ and $\delta\chi_{\nu_i,1}$ are the uncertainties in the mass of the state with $S=1$ and in the zero of Bessel function that is related to the mass $m_i$. The uncertainty $\delta m_0$ of the state with $S=1$ for each meson family in our model will be the computed as follows: we consider $\delta m_0$ in the family $\rho(770)$ as the arithmetic mean of all uncertainties in the PDG data for this family, which represents 1.3\% of the mass of $\rho(770)$. In order to have a uniform criterium, we will take the uncertainties $\delta m_0$ of each family to be 1.3\% of their $S=1$ state mass.

\begin{table}[t!]
\begin{tabular}{|c|c|c|c|c|}
\hline
\hline 
$I^{G}(J^{PC})$ & Meson & $a_{\rm Log}$  & AHW (Log) Regge Trajectory (GeV) &$R^2$
\\ \hline
 $1^{+}(1^{--})$ & $\rho (770)$ &    
$2.77\pm0.12$ & 
 $J=(0.88\pm0.01)M^2 +0.50\pm0.03$ &0.9998
\\
\hline
 $0^{-}(1^{--})$ & $\omega (782)$ &  
$2.61\pm0.09$ &
 $J=(0.90\pm0.01)M^2 +0.48\pm0.03$ &0.9997
\\ \hline
 $1^{+}(1^{+-})$& $b_1 (1235)$ & 
$0.56\pm0.03$ &
 $J=(0.78\pm0.02)M^2-0.13\pm0.08$ &0.9985
\\ \hline
 $0^{-}(1^{+-})$&$h_1 (1170)$ & 
$0.95\pm0.06$ & 
 $ J=(0.73\pm0.01)M^2+0.04\pm0.05$ & 
0.9993
\\
 \hline 
\end{tabular}
\caption{\label{Tlog} The Regge trajectories in GeV units for each light unflavored meson family (same $PC$ parity and same isospin $I$) from the AHW model, Eqs. \eqref{anomdim} and \eqref{masstower2}, with the corresponding values of $R^2$ from Eq.  \eqref{Rquadrado}. The coefficients $a_{\rm Log}$ presented here come from the minimization of $Q$, Eq. \eqref{minimization}.}
\end{table}

Now that we have defined our criteria  for best fit and how we compute the uncertainties  $\delta{m_i}$ for each calculated mass $m_i$, we are able to tackle the AHW model for light unflavored mesons.


\begin{figure}[t!]
\vskip0.5cm 
\centering
\includegraphics[width=7.5cm]{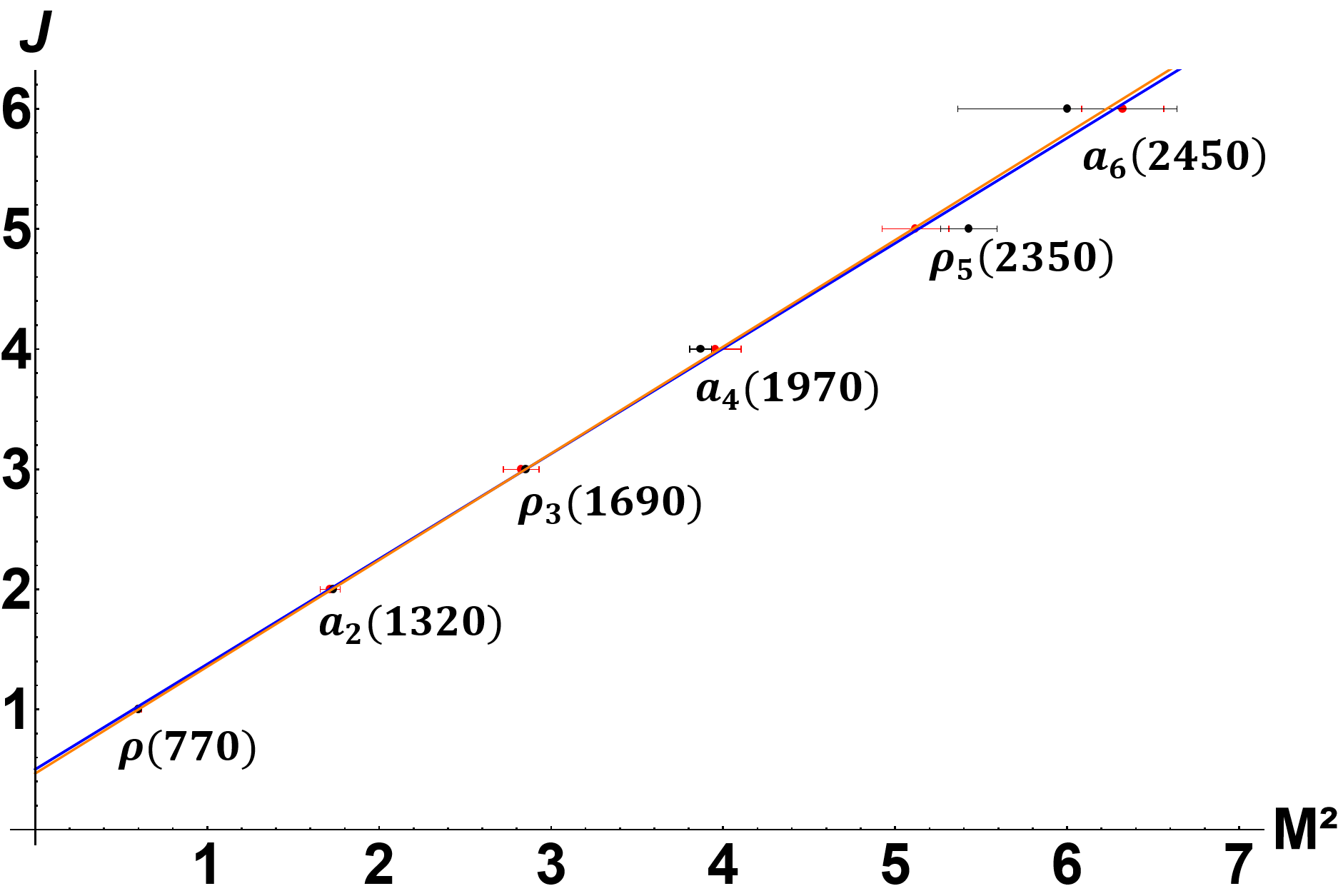}\qquad 
\includegraphics[width=7.5cm]{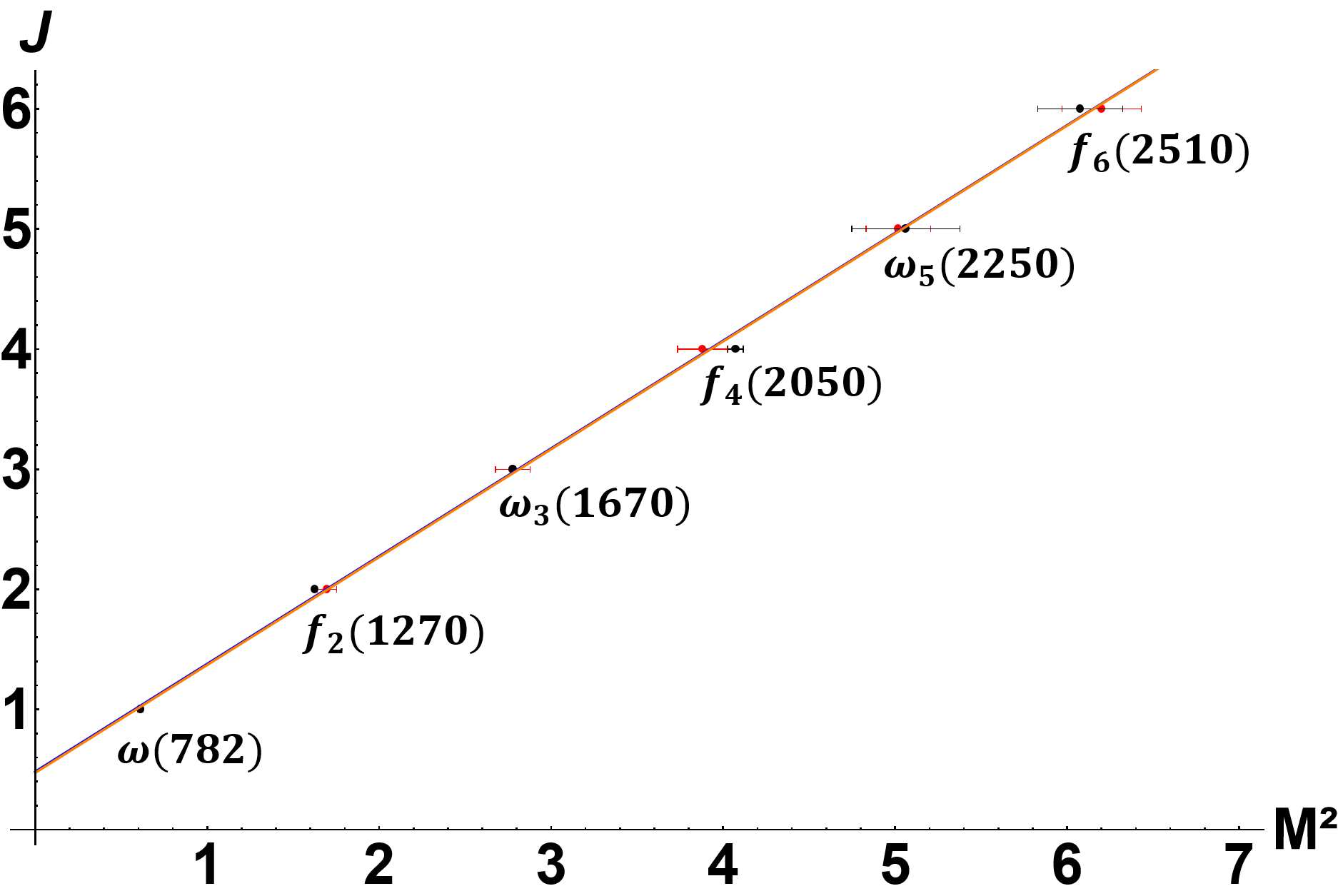}\vskip 0.5cm 
\includegraphics[width=7.5cm]{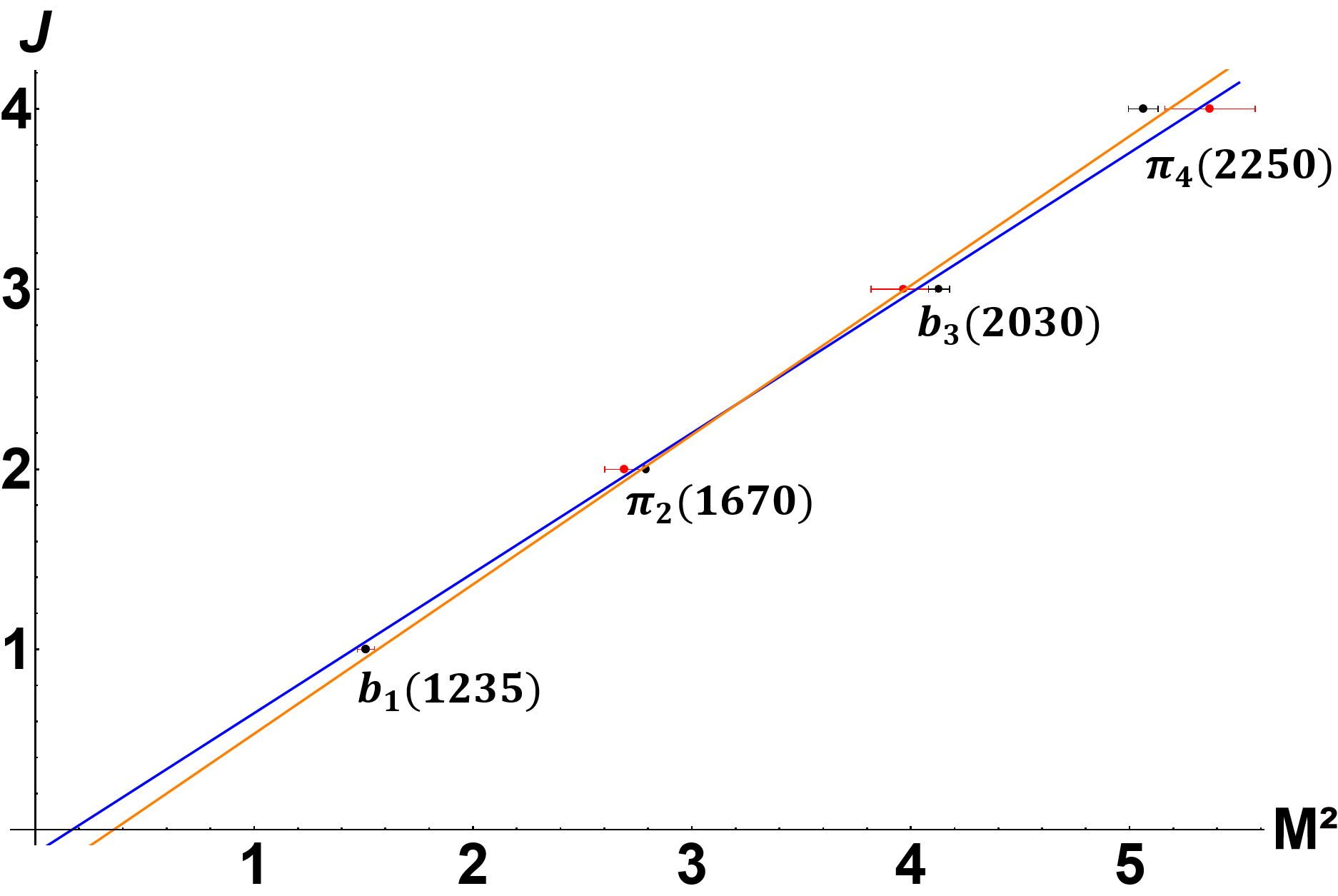}\qquad 
\includegraphics[width=7.5cm]{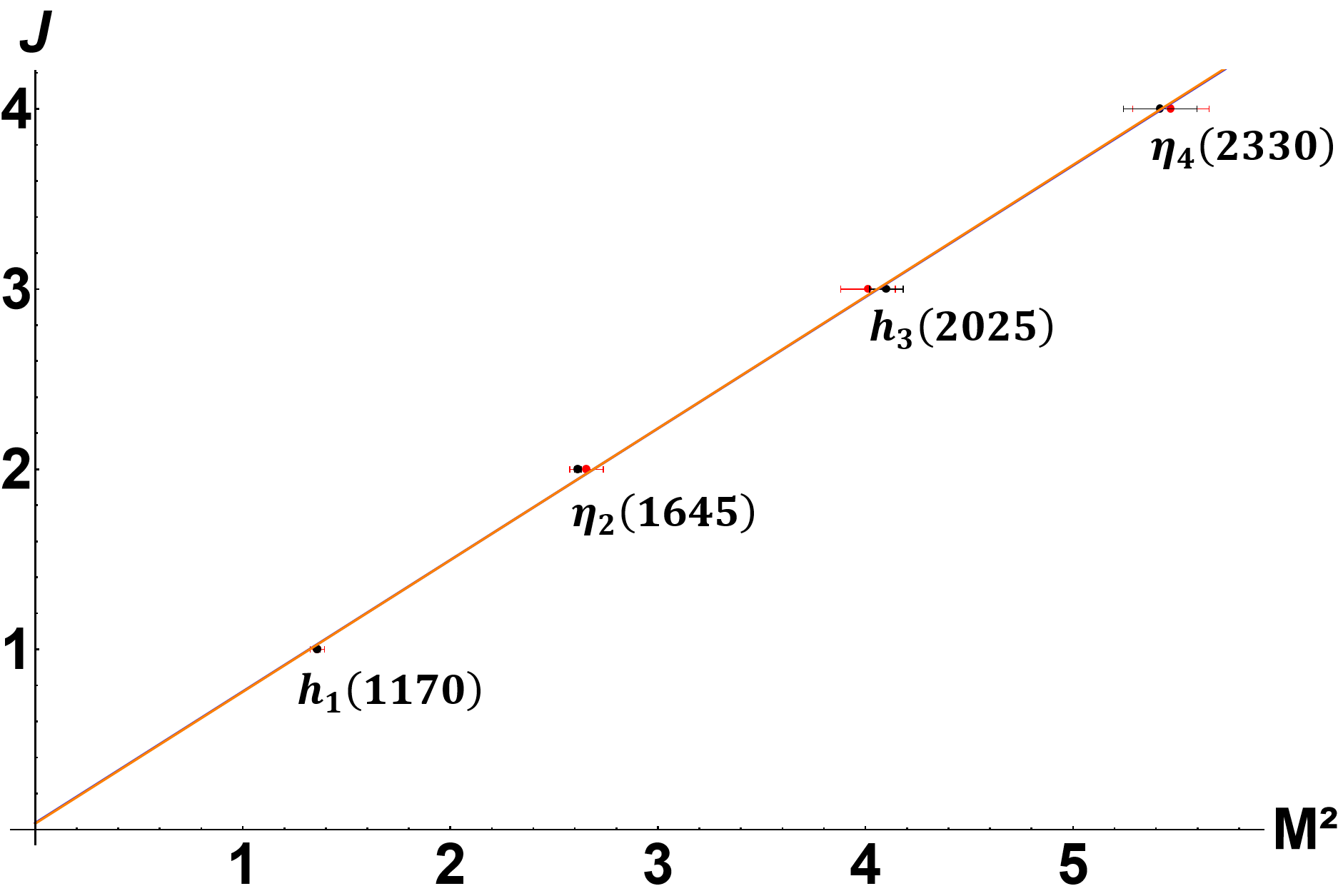}
\caption{Regge trajectories ($J\times M^2$) for mesons, with masses $M$ expressed in GeV, from the AHW Log model (blue lines) using the minimization of $Q$, Eq. \eqref{minimization}, and from PDG data (orange lines). Red dots are the values obtained from Eq. \eqref{masstower2}, while black dots from PDG data, both shown with uncertainties in Tables  \ref{HWlog} and \ref{tabpdg}, respectively. The lines are  obtained by linear regression of these points. {\sl Upper left panel:} mesons with even ${PC}$, $I=1$ ($\rho (770)$ family); {\sl Upper right panel:}  mesons with even ${PC}$, $I=0$ ($\omega (782)$ family); {\sl Lower left panel:}  mesons with odd ${PC}$, $I=1$ ($b_1 (1235)$ family); {\sl Lower right panel:}  mesons with odd ${PC}$, $I=0$ ($h_1 (1170)$ family).}
\label{flog}
\end{figure}

    
    We start applying the AHW model for mesons with even ${PC}$, isospin $I=1$, and spins $S\ge1$, taking the mass of the  $\rho (770)$ state from PDG as an input, which fixes $\Lambda_{\rm HW}=150.96$ MeV for this family. We minimize the parameter $Q$, given by Eq. \eqref{minimization}, with respect to PDG masses from  Table \ref{tabpdg}, finding the logarithm anomalous dimensions, Eq. (\ref{anomdim}), to be 
\begin{equation}\label{anomdimrho}
        \Delta_{\rm Log}(S)=(2.77\pm0.12)\ln(S).
    \end{equation}
     The masses with their  uncertainties obtained within this model, from Eq. \eqref{masstower2}, with these  anomalous dimensions, are presented in Table \ref{HWlog}, together with percentage deviations in comparison with PDG data. The value of $\Delta_{\rm Log} (S)$ of the above equation for each spin $S$ is also shown in the last column of this Table. 
The Regge trajectory obtained by doing linear regression with the obtained masses from this AHW Log model is given by
    \begin{equation}\label{regge770log}
        J=(0.88\pm0.01)M^2+0.50\pm0.03. 
    \end{equation}
    This trajectory is presented in the upper left panel of Fig. \ref{flog} as the blue line, together with the one from PDG (orange line). The red dots are the values in the $J\times M^2$ plane of the mesons $\rho(770)$, $a_2(1320)$, $\rho_3(1690)$, $a_4(1970)$, $\rho_5(2350)$, and $a_6(2450)$ 
    in this AHW Log model, with their uncertainties, while the black dots represent PDG data from Table \ref{tabpdg}.  

    The above analysis presented for the $\rho(770)$ meson family is also applied for the other meson families, namely $\omega(782)$, $b_1 (1235)$, and $h_1 (1170)$ which results are shown in Tables \ref{HWlog} and  \ref{Tlog}, and in the upper right and lower left and right panels of Figure \ref{flog}, respectively. These results are very good when compared with PDG data, Table \ref{tabpdg}, and present percentage deviations of less than 3$\%$.

For completeness and comparison, we also present in the Appendix \ref{Appendix} a more standard minimization of $\chi^2$, defined in Eq. \eqref{chiquadrado}. As one can check, the results of both minimizations are very close to each other (actually, they coincide for the $h_1(1170)$ family, within the precision we are working here).







\section{Linear AHW for mesons}\label{LHW}

    Recently, we have shown \cite{Costa-Silva:2023vuu} that one can obtain asymptotically linear Regge trajectories for glueballs with even spins $S$ from an AHW model, when considering a total dimension, $\Delta_{\rm Total}$, proportional to $\sqrt{S}$. By asymptotically linear Regge trajectories, we mean functional relations between $J$ and $M^2$ which are not exactly linear but can approximate very well the linear behavior with the desired precision.

    In order to find asymptotically linear Regge trajectories for light unflavored mesons with spins $S\ge1$, we apply a similar procedure here with a small modification in the power of the spins $S$ as
    \begin{equation}\label{lineardim}
        \Delta_{\rm LAHW}=a_{\rm LAHW}(S^{\frac{1}{2}+ \epsilon}-1)+4,
    \end{equation}
    where in the ideal case, $\epsilon\rightarrow 0$. Note that when $S=1$, which is the lowest spin state for each mesonic family considered here, we recover the canonical dimension $\Delta=4$, compatible with Eq. \eqref{canonicaldim}. 

  \begin{table*}[ht!]
\begin{tabular}{|c|c|c|c|c|c|c|}
\hline 
\hline 
$I^{G}(J^{PC})$ & Meson & Linear AHW & $\delta_{\rm LAHW}$&$\Delta_{\rm LAHW}$ 
\\ \hline
 $1^{+}(1^{--})$ & $\rho (770)$ & $775.26\pm10.08$&$0\%$&$0$ 
  \\
 $1^{-}(2^{++})$&$a_2 (1320)$&$1296.75\pm21$&$1.53\%$&$1.85$
 \\
 $1^{+}(3^{--})$& $\rho_3 (1690)$&$1678.61\pm30$&$0.60\%$&$3.03$
 \\
 $1^{-}(4^{++})$ & $a_4 (1970)$&$1993.84\pm38$&$1.36\%$&$3.87$
 \\
 $1^{+}(5^{--})$& $\rho_5 (2350)$ &$2268.03\pm44$&$2.66\%$&$4.49$
 \\
 $1^{-}(6^{++})$& $a_6 (2450)$&$2513.71\pm51$&$2.60\%$&$4.96$
\\ \hline
 $0^{-}(1^{--})$ & $\omega (782)$& $782.66\pm10.17$&$0\%$&$0$
  \\
 $0^{+}(2^{++})$ & $f_2 (1270)$&$1294.28\pm20$&$1.47\%$&$1.76$
 \\
 $0^{-}(3^{--})$& $\omega_3 (1670)$&$1669.09\pm29$&$0.13\%$&$2.88$
 \\
 $0^{+}(4^{++})$& $f_4 (2050)$&$1978.53\pm37$&$1.96\%$&$3.67$
 \\
 $0^{-}(5^{--})$& $\omega_5 (2250)$&$2247.67\pm44$&$0.10\%$&$4.24$
 \\
 $0^{+}(6^{++})$& $f_6 (2510)$&$2488.81\pm50$&$0.97\%$&$4.67$
\\ 
\hline
 $1^{+}(1^{+-})$& $b_1 (1235)$&$1229.5\pm16$&$0\%$&$0$
 \\
 $1^{-}(2^{-+})$& $\pi_2 (1670)$&$1667.30\pm26$&$0.20\%$&$0.48$
 \\
 $1^{+}(3^{+-})$& $b_3 (2030)$&$2000.60\pm37$&$1.55\%$&$0.65$
 \\
 $1^{-}(4^{-+})$& $\pi_4 (2250)$&$2281.03\pm46$&$1.38\%$&$0.65$
\\ 
\hline
 $0^{-}(1^{+-})$& $h_1 (1170)$&$1166\pm15$&$0\%$&$0$
\\
 $0^{+}(2^{-+})$&$\eta_2 (1645)$&$1646.87\pm26$&$1.85\%$&$0.72$
 \\
 $0^{-}(3^{+-})$ & $h_3 (2025)$&$2009.96\pm37$&$0.74\%$&$1.07$
 \\
 $0^{+}(4^{-+})$ & $\eta_4 (2330)$&$2314.19\pm47$&$0.59\%$&$1.22$
 \\
 \hline 
 \hline 
\end{tabular}
\caption{\label{linearT} Masses in MeV for mesons  classified under even or odd $PC$ with isospins $I=0,1$ and spins $S\ge1$, from the Linear AHW model, determined from Eq. \eqref{masstower3}, and the percentage deviations $\delta_{\rm LAHW}=\delta_{i}\times 100\%$, with $\delta_{i}$ given by Eq. \eqref{percentualdev},  of the masses from this model compared with PDG data,  using  minimization with respect to the  parameter $Q$, Eq. \eqref{minimization}. For each family the mass of the state with $J=S=1$  is taken as an input. In the last column we show the total  dimension $\Delta_{\rm LAHW}$ for each state determined by Eq. \eqref{lineardim}.}
\end{table*}

\begin{table}[t!]
\footnotesize 
\begin{tabular}{|c|c|c|c|c|c|c|}
\hline
\hline 
$I^{G}(J^{PC})$ & Meson & $a_{\rm LAHW}$ & $\epsilon$ & Linear AHW  Regge Trajectory (GeV) & $n$&$R^2$
\\ \hline
  $1^{+}(1^{--})$& $\rho (770)$ &    
$6.87\pm0.16$ &   
$0.00\pm0.005$ & 
 $J=(0.87\pm0.01)M^2 +0.51\pm0.02$ 
 
 & 6&0.999793
 \\
   & &  &  &  
 $J=(0.8479\pm0.0002)M^2 +0.60\pm0.01$
 & 50 &0.999997
 \\
\hline
 $0^{-}(1^{--})$ & $\omega (782)$ &  
$6.67\pm 0.13$ &
$0.00\pm0.005$ & 
 $J=(0.89\pm0.01)M^2 +0.49\pm0.02$ 
 
 & 6&0.999845
\\ 
  & & & & 
 $J=(0.8766\pm0.0003)M^2 +0.52\pm0.01$ 
 & 50&0.999995
 \\
\hline
 $1^{+}(1^{+-})$& $b_1 (1235)$ & 
$3.19\pm 0.03$ &
 $0.05\pm0.005$ & 
 $J=(0.81\pm0.01)M^2-0.24\pm0.03$ 
 & 4&0.999846
 \\
  & & & &  
 $J=(0.8549\pm0.0003)M^2-0.42\pm0.01$ 
 & 50 &0.999993
\\ \hline
 $0^{-}(1^{+-})$& $h_1 (1170)$ & 
$3.79\pm 0.08$ & 
 $0.04\pm0.005$ & 
 $ J=(0.75\pm0.03)M^2
 -0.03\pm0.01$ 
 & 4&0.999960
 \\
  & & & &  
 $ J=(0.7667\pm0.0003)M^2
 -0.006\pm0.01$ 
 & 50 &0.999994
 \\
 \hline 
\end{tabular}
\caption{\label{Tlin} The Regge trajectories in GeV units for each light unflavored meson family (same $PC$ parity and same isospin $I$) from the Linear AHW model, Eq. \eqref{masstower3}. For each meson family we show two Regge trajectories corresponding to the linear regression of two sets of points with dimension $n$. These trajectories are plot in Figures \ref{linearfig} and \ref{linearfig50}. We also show the corresponding values of  $R^2$ from Eq.  \eqref{Rquadrado}. The coefficients $a_{\rm LAHW}$ and $\epsilon$ presented here come from the minimization of $Q$, Eq. \eqref{minimization}.}
\end{table}

    Here, the parameter $a_{\rm LAHW}$ is obtained exactly as in the AHW Log model, by minimizing $Q$ in Eq. (\ref{minimization}). The parameter $\epsilon$ that represents a little deviation of the ideal case $\epsilon=0$ is chosen to obtain the best asymptotically linear Regge trajectory for each meson family. The value of $\epsilon$ is  the one that minimizes the total deviation of the masses from the straight line representing the Regge trajectory.

    The masses that follows from this linear AHW model are determined by 
\begin{equation}\label{masstower3}
        m_{\nu,1} = \chi_{\nu,1} \Lambda_{\rm HW}; \qquad\qquad (\nu=\Delta_{\rm LAHW}-2),  
    \end{equation}
    where $\Delta_{\rm LAHW}$ is defined in Eq. \eqref{lineardim}. 
Since $\Delta_{\rm LAHW}=4$ for $S=1$, the values of $\Lambda_{\rm HW}$ here in this model, for each meson family, are the same as those in the AHW Log model, as well as in the original HW presented in Table \ref{HWSW}.

   As in the case of logarithmic anomalous dimension, we will analyze the families of light unflavored mesons 
   with even or odd $PC$, isospins $I=0,1$, and spins $S\ge1$. 
 The Regge trajectories that we obtain within this model are asymptotically linear for each meson family with the appropriate choice of parameters. 

 \begin{figure}[t!]
\vskip0.5cm 
\centering
\includegraphics[width=7.5cm]{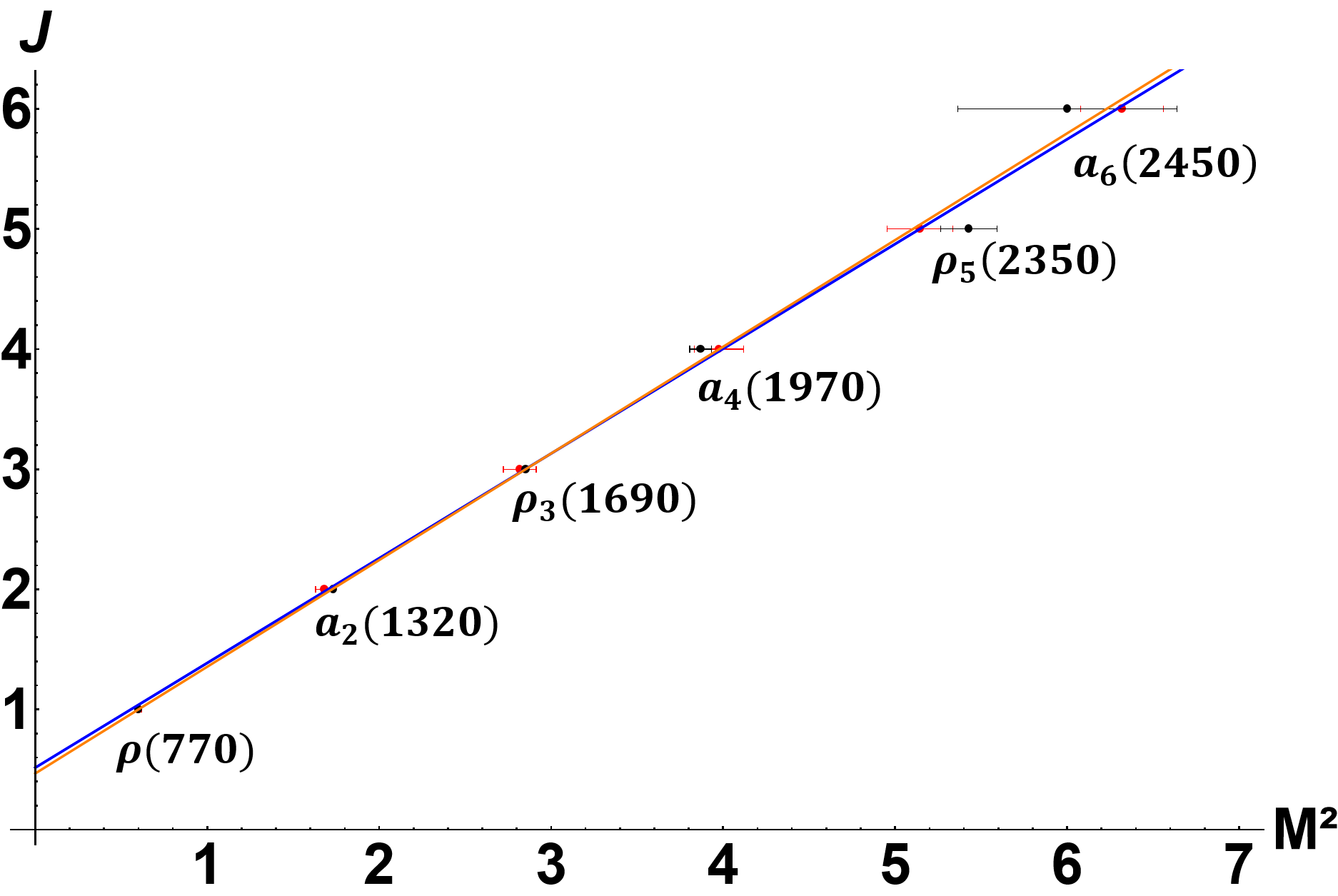}\qquad 
\includegraphics[width=7.5cm]{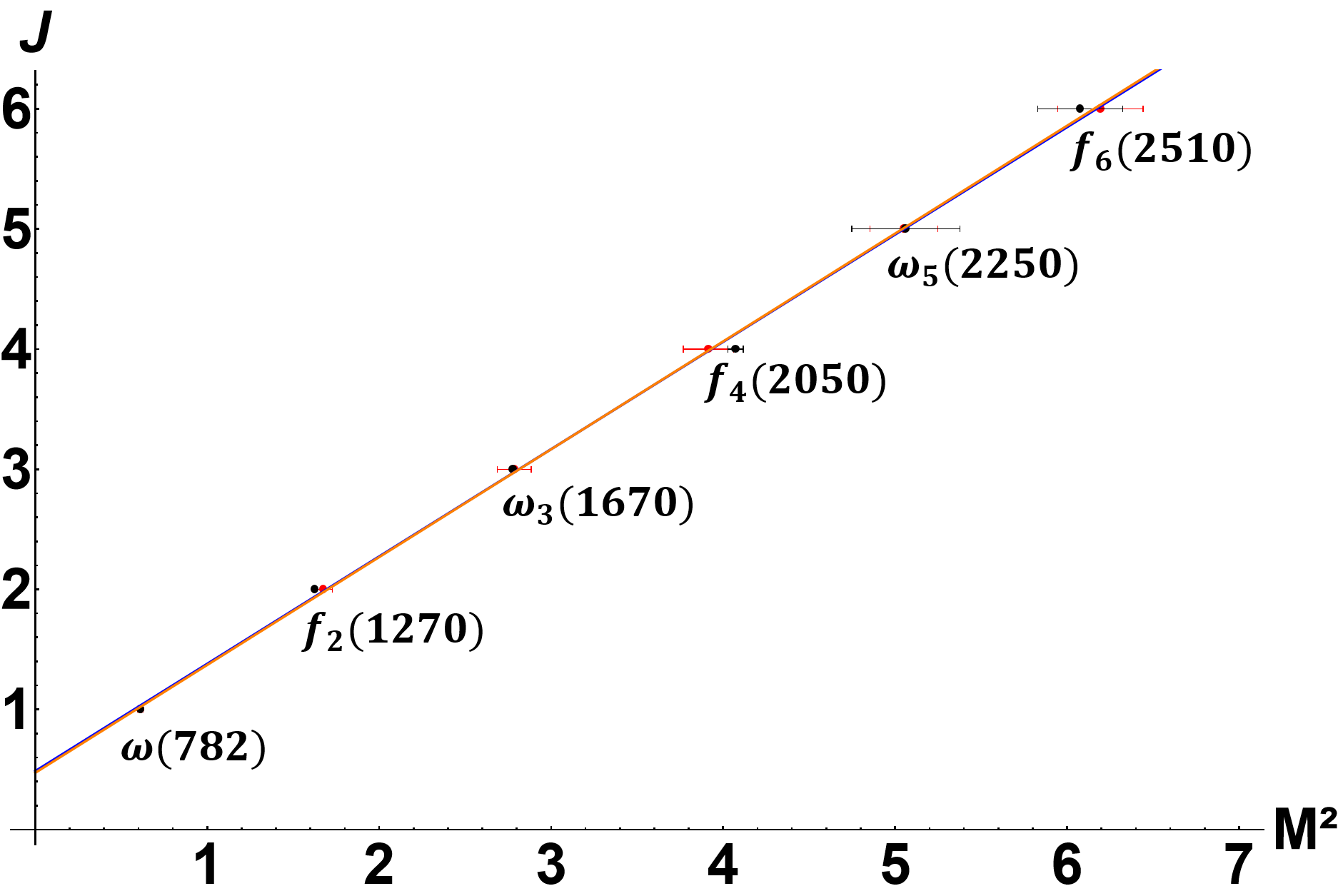}\vskip 0.5cm 
\includegraphics[width=7.5cm]{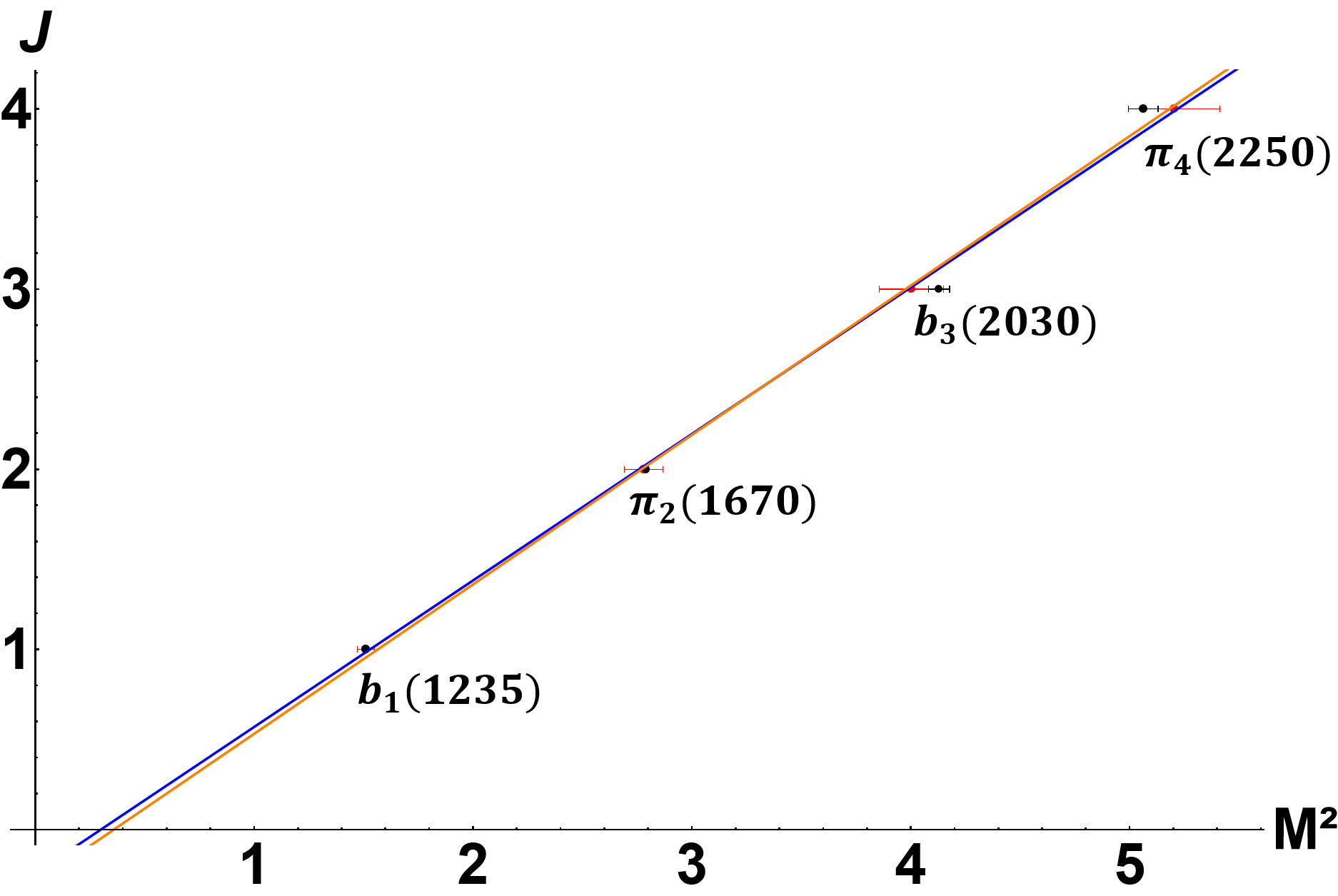}\qquad 
\includegraphics[width=7.5cm]{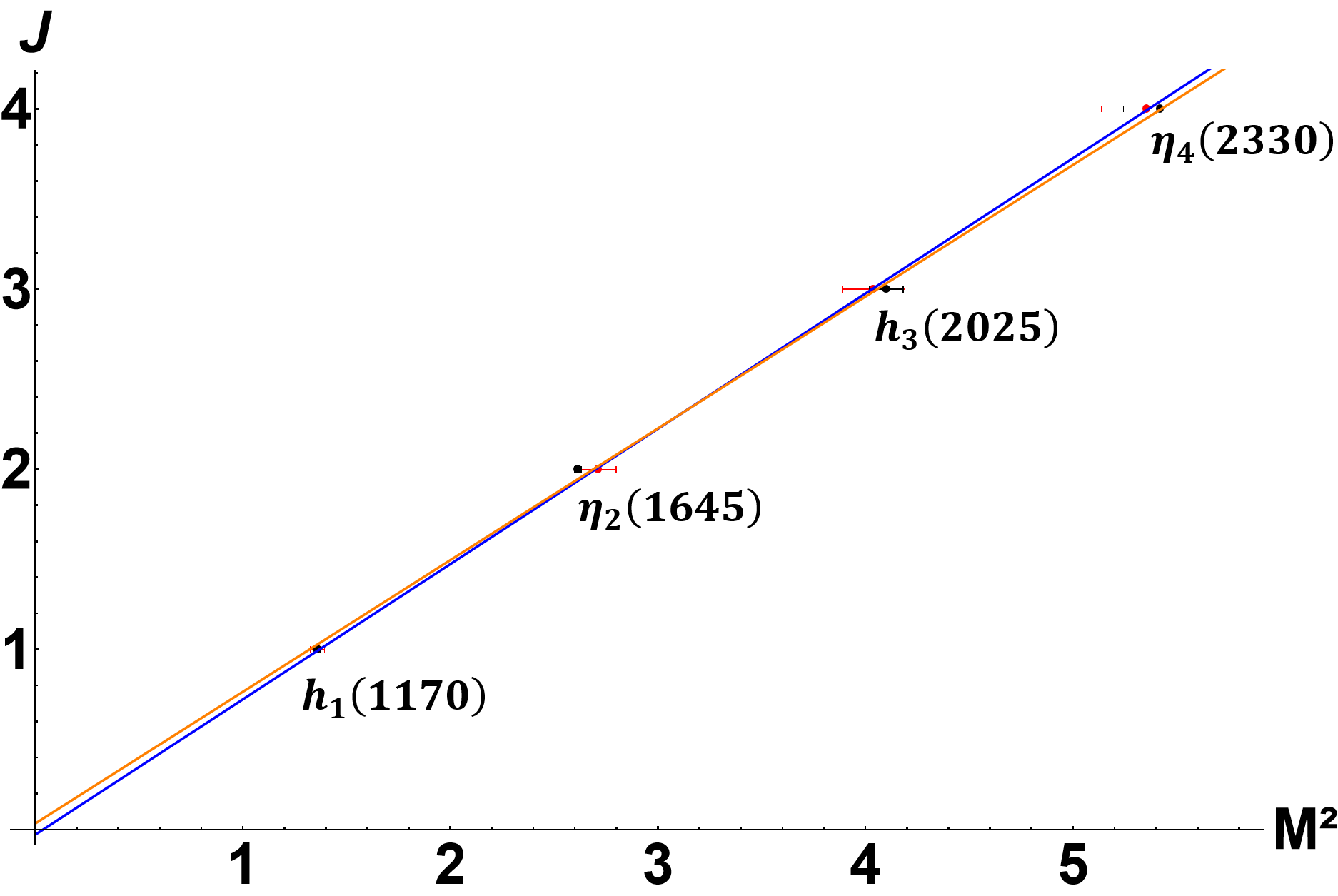}
\caption{Regge trajectories ($J\times M^2$) for mesons, with masses $M$ expressed in GeV, from the Linear AHW model (blue lines) with total dimension given by Eq. \eqref{lineardim},  using the minimization of $Q$, Eq. \eqref{minimization}, and from PDG data (orange lines). Red dots are the values obtained from Eq. \eqref{masstower3}, while black dots come from PDG, both with uncertainties. The trajectories are obtained by linear regression of these points, for each meson family. {\sl Upper left panel:} mesons with even $PC$, $I=1$ ($\rho (770)$ family); {\sl Upper right panel:}  mesons with even ${PC}$, $I=0$ ($\omega (782)$ family); {\sl Lower left panel:}  mesons with odd ${PC}$, $I=1$ ($b_1 (1235)$ family); {\sl Lower right panel:}  mesons with odd ${PC}$, $I=0$ ($h_1 (1170)$ family).}
\label{linearfig}
\end{figure}

\begin{figure}[t!]
\vskip0.5cm 
\centering
\includegraphics[width=7.5cm]{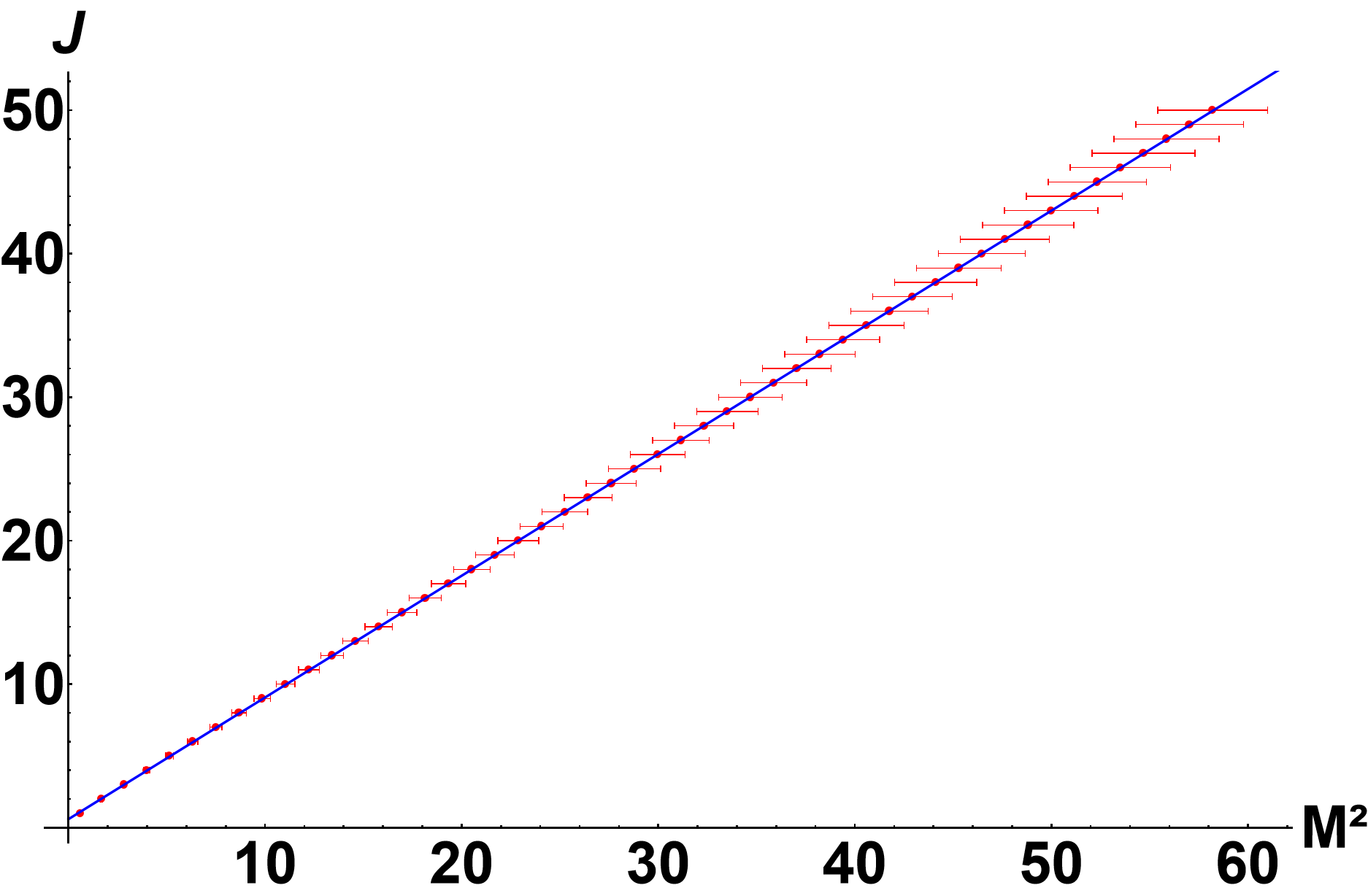}\qquad 
\includegraphics[width=7.5cm]{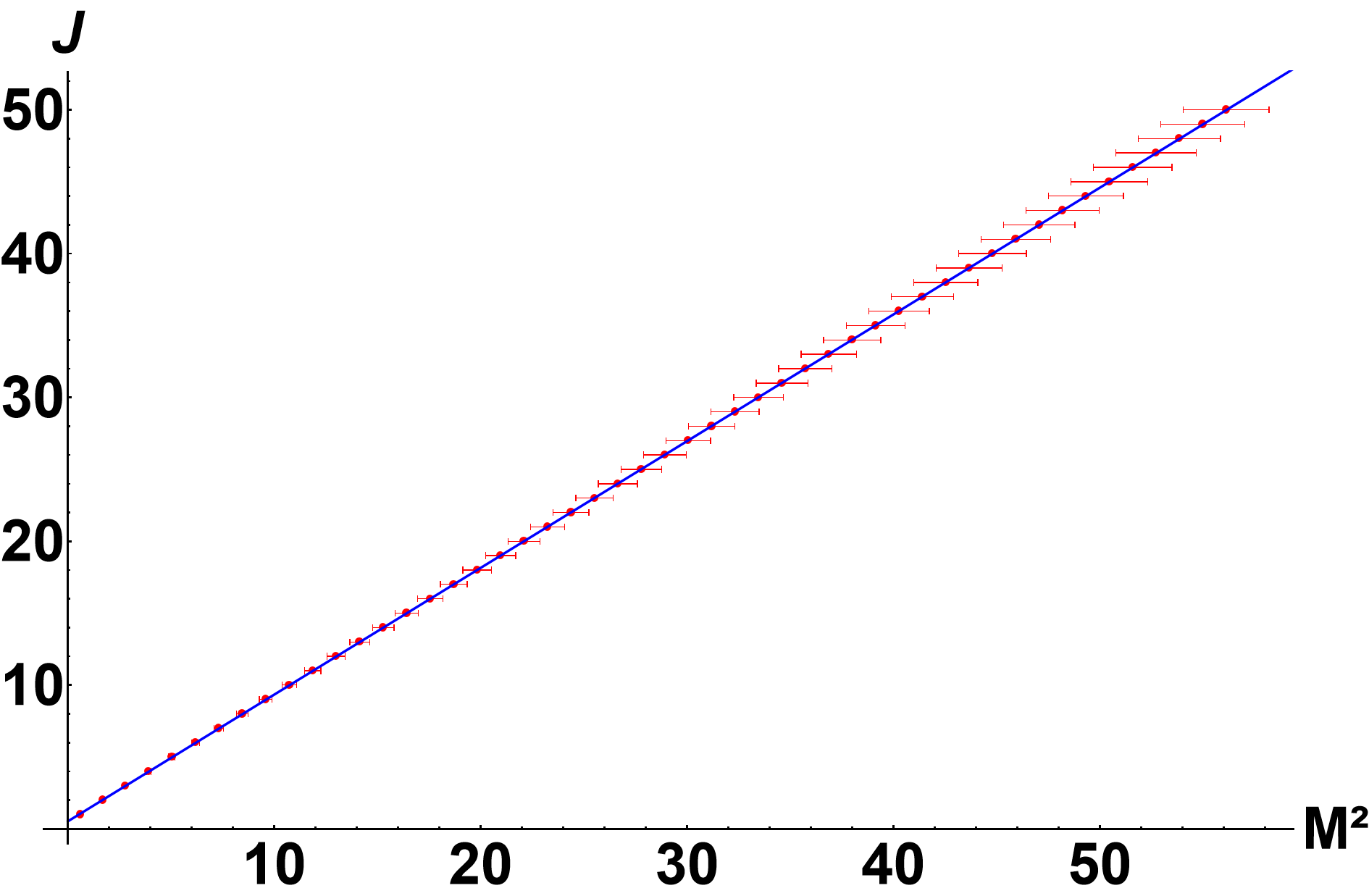}\vskip 0.5cm 
\includegraphics[width=7.5cm]{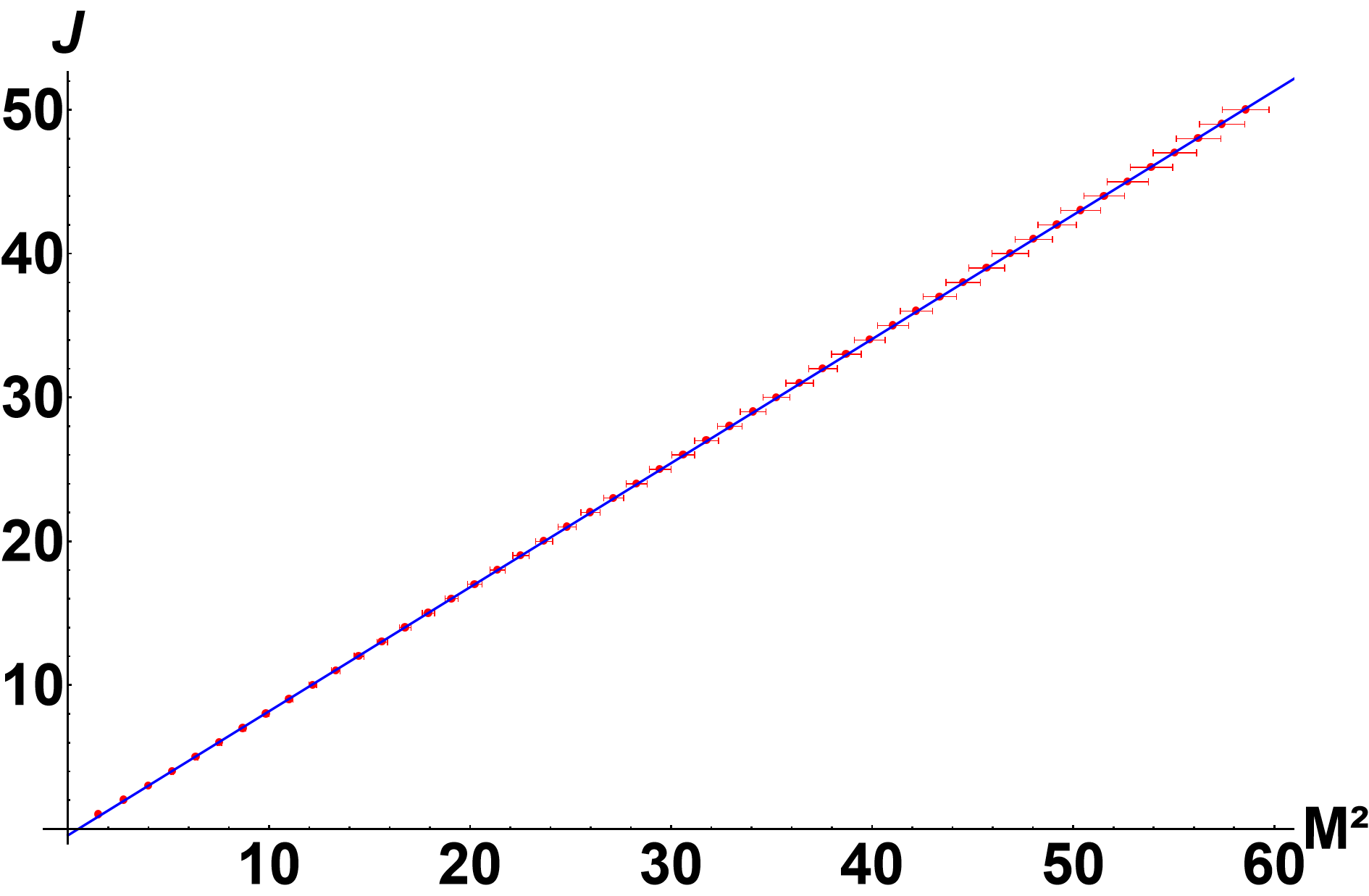}\qquad 
\includegraphics[width=7.5cm]{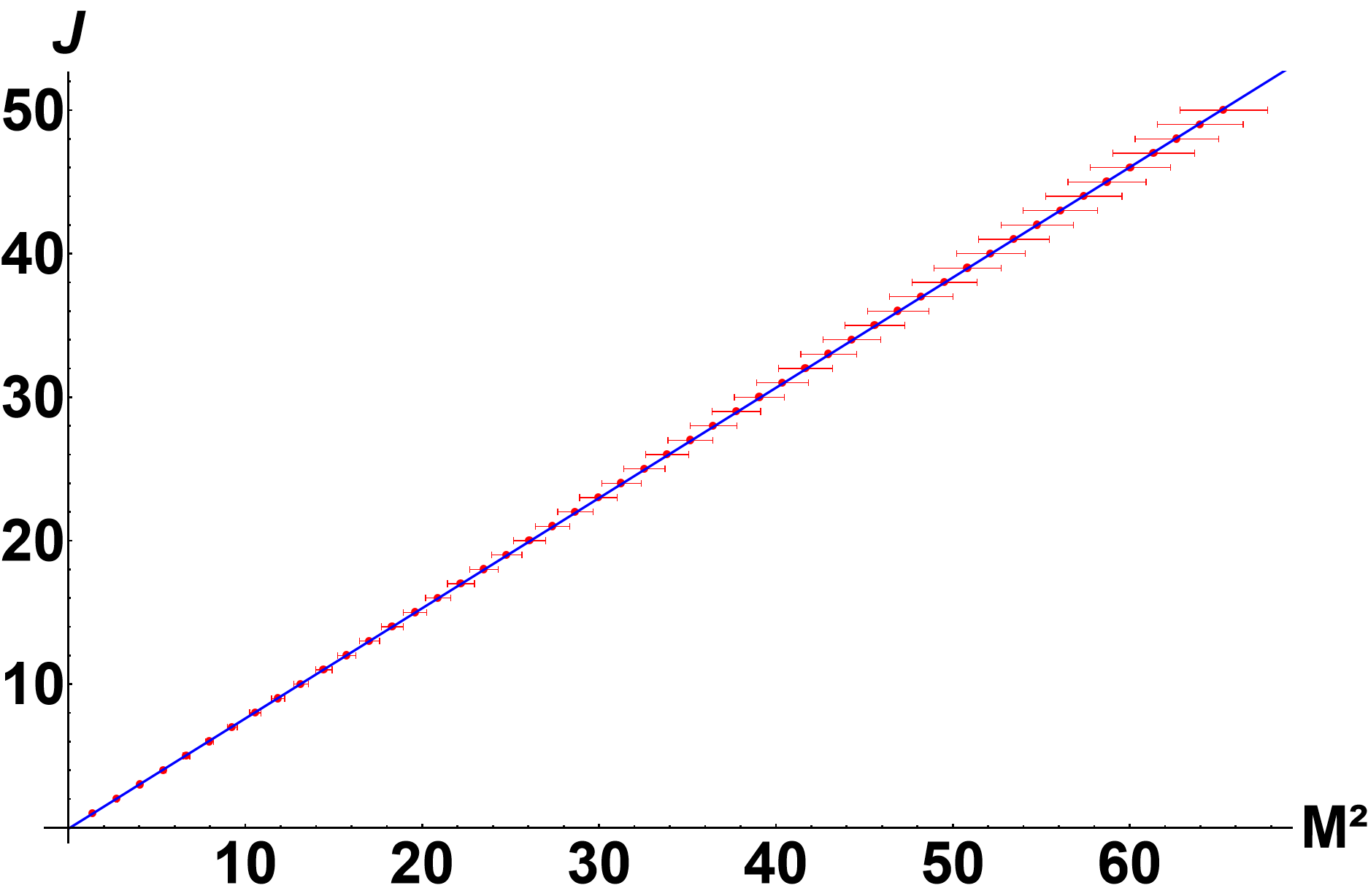}
\caption{Regge trajectories ($J\times M^2$) for mesons, with masses $M$ expressed in GeV, from the linear AHW model, with total dimension given by Eq. \eqref{lineardim}, using the minimization of $Q$, Eq. \eqref{minimization}. Red dots are the values obtained from Eq. \eqref{masstower3} with uncertainties and the blue lines are the Regge trajectories, obtained by linear regression of these 50 points, for each meson family. {\sl Upper left panel:} mesons with even ${PC}$, $I=1$ ($\rho (770)$ family); {\sl Upper right panel:}  mesons with even ${PC}$, $I=0$ ($\omega (782)$ family); {\sl Lower left panel:}  mesons with odd ${PC}$, $I=1$ ($b_1 (1235)$ family); {\sl Lower right panel:}  mesons with odd ${PC}$, $I=0$ ($h_1 (1170)$ family).}
\label{linearfig50}
\end{figure}


As in the previous section, we start with the $\rho(770)$ family, {\sl i.e.}, for mesons with even ${PC}$, isospin  $I=1$, and spin $S\ge1$, the parameters in Eq. (\ref{lineardim}) that minimize $Q$ are $a_{\rm LAHW}=6.87\pm0.16$ and $\epsilon=0.00\pm0.005$, so the total dimension reads
    \begin{equation}
        \Delta_{\rm LAHW}=(6.87\pm0.16)\sqrt{S}-2.87\pm0.16.
    \end{equation}
    The masses obtained here, using Eq. \eqref{masstower3}, are presented in Table \ref{linearT}, with respectives uncertainties, using the mass of the state $\rho(770)$ as an input from PDG. The Regge trajectory for these mesons is given by
    \begin{equation}\label{RTI=1even}
        J=(0.87\pm0.01)M^2+0.51\pm0.02,
    \end{equation}
     and is presented in the upper left panel of Fig. \ref{linearfig} as the blue line, together with the PDG data represented by black dots with Regge trajectory given by the orange line. The red dots represent the masses of the LAHW model with their uncertainties.

    In order to show that the linear AHW model indeed presents asymptotically linear Regge trajectories we also  plot $J\times M^2$ for these mesons  with  even $PC$, isospin $I=1$, and $S\ge1$, for the first 50 states with masses calculated from this model. The trajectory in this case becomes
    \begin{equation}\label{RTI=1even50}
    J=(0.8479\pm0.0002)M^2+0.60\pm0.01.
    \end{equation}
    
    This trajectory is presented in the upper left panel of Fig.~\ref{linearfig50}. The  trajectories from Eqs. \eqref{RTI=1even} and \eqref{RTI=1even50} 
    can be compared with the one obtained from PDG data shown in Table \ref{tabpdg} and are in good agreement with it.
These results are synthesized in Table \ref{Tlin}.  

This analysis is also employed for the other meson families studied here, $\omega(782)$, $b_1(1235)$, and $h_1(1170)$, which results are collected in Tables  \ref{linearT} and \ref{Tlin}, and the other panels of Figs. \ref{linearfig} and \ref{linearfig50}. As one can see from these Figures, the agreement between the predictions of the LAHW model and PDG data is very good. This conclusion is also supported by an inspection of Table  \ref{linearT} which presents percentage deviations with respect to PDG data of less than 3$\%$. The Regge trajectories presented in Table \ref{Tlin} are also in very good agreement with the ones from PDG, shown in Table \ref{tabpdg}. 

     For comparison, we also present in the Appendix \ref{Appendix} the Tables \ref{linearTchi2} and \ref{Tlinchi2}, with the results of the minimization of $\chi^2$ applied to the LAHW model.  As one can see from these tables, the results obtained from this method are very close to the minimization of $Q$, Eq. \eqref{minimization}, within this model, shown in Tables  \ref{linearT} and \ref{Tlin}.



\section{Discusssion and Conclusions}
\label{conclusions}

In this work, we proposed anomalous and linear HW models for light unflavored mesons. We have shown that the masses and Regge trajectories produced by these models are better than those  obtained from the original HW and SW models, shown in Table \ref{HWSW} (see, {\sl e.g.}, \cite{Boschi-Filho:2002xih, Boschi-Filho:2002wdj, deTeramond:2005su, Erlich:2005qh, Boschi-Filho:2005xct, Karch:2006pv, Colangelo:2008us}). In fact, it is remarkable that the mass percentage deviations predicted by the AHW and LAHW models proposed here for mesons are less than $3\%$ when compared with PDG data. 

In particular, the linear AHW model produce asymptotically linear Regge trajectories, circumventing the well-kown problem of non-linearity of the original HW model.

Since we analyzed four meson families classified under $PC$ parity (even or odd) and isospin $I$ values (0 or 1) starting with the states  $\rho (770)$ (even $PC$, $I=1$), $\omega (782)$ (even $PC$ and $I=0$), $b_1 (1235)$ (odd $PC$ and $I=1$), and $h_1 (1170)$ (odd $PC$ and $I=0$) with the AHW and LAHW models it is possible to discuss some similarities and differences between these families. First of all, the similarities of the $\rho (770)$ and  $\omega (782)$ families are very clear from the comparison of the parameters found by best fit to describe them. For instance, looking at Table \ref{Tlog} one sees that the coefficients $a_{\rm Log}$ for these two families are close to each other. The same can be said about the Regge trajectories equations appearing in this table. The same observation holds when analyzing Table \ref{Tlin}, with respect to the parameters $a_{\rm LAHW}$ and $\epsilon$, and the corresponding trajectories. Since these two families share the same even $PC$ parity but different isospins, one can conclude that the dependence on the isospin of these parameters is small. The same conclusion is reached with the families of the $b_1 (1235)$ and $h_1 (1170)$ mesons which also have the same odd $PC$ parity, but different isospins. 

This discussion could also help us understand the role of the $\epsilon$ parameter in Eq. \eqref{lineardim}. Since we found $\epsilon=0$ for mesons with even $PC$ parity, while $\epsilon\sim 0.05$ in the case of odd $PC$ parity, it seems that this parameter is sensitive to these quantum numbers. This fact is related to the coefficients $\alpha'$ and $\alpha_0$ of the linear Regge trajectories ($J=\alpha'M^2 + \alpha_0$) found for these mesonic families, in the linear anomalous HW model. In the case of even $PC$ parity, $\epsilon=0$,  $\alpha'\sim 0.9 \,{\rm GeV}^{-2}$ and $\alpha_0\sim 0.50$, while for odd $PC$ parity one has $\epsilon\sim 0.05$,  $\alpha'\sim 0.8 \,{\rm GeV}^{-2}$ and $\alpha_0 <0$. From the physical point of view, one can say that the energy necessary to create a meson state with a given spin $S=J$~ is lower for an even $PC$ parity one than the case with odd $PC$ parity with the same spin. This is in agreement with the experimental data from PDG, as can be seen in Table~\ref{tabpdg}. Then, it is clear that the nonzero values of $\epsilon$ are related to an extra energy interaction required to form a mesonic state with odd $PC$ parity. 

In the case of the AHW Log model, an analogous discussion might be made looking at the values of the $a_{\rm Log}$ parameter, which is $\sim 2.7$ for the even parity case, while $\sim 0.75$ for the odd case. So one could speculate a relation like $a_{\rm Log}\sim 0.75 + 2\delta_{PC}$, where $\delta_{PC}$ is zero in the odd case ($+-$ or $-+$) and 1 in the even case ($++$ or $--$). A similar relation can also be proposed for the LAHW model with $a_{\rm LAHW}\sim 3.5 + 3.3\,\delta_{PC}= 6.8 - 3.3\,\delta_{PC}$, which could be written more precisely in terms of the parameter $\epsilon$ as $a_{\rm LAHW}\sim 6.8 -  72\,\epsilon$. 

Finally, we expect to successfully extend the analysis presented here to other particles such as heavy and heavy-light mesons, as well as baryons. This is presently under investigation. 

\section*{Acknowledgments} 
 R.A.C.-S. is supported by  Conselho Nacional de Desenvolvimento Científico e Tecnológico (CNPq) and Coordenação de Aperfeiçoamento de Pessoal de Nível Superior (CAPES) under finance code 001. H.B.-F. is partially supported by Conselho Nacional de Desenvolvimento Cient\'{\i}fico e Tecnol\'{o}gico (CNPq) under grant $\#$ 310346/2023-1.

\section{Appendix: $\chi^2$ minimization}\label{Appendix}

\begin{table*}[ht!]
\begin{tabular}{|c|c|c|c|c|}
\hline 
\hline 
$I^{G}(J^{PC})$ & Meson &  AHW Log ($\chi^2$)&$\delta_{\rm Log}$&$\Delta_{\rm Log}$
\\ \hline
 $1^{+}(1^{--})$ & $\rho (770)$&$775.26\pm10.08$&$0\%$&$0$
 \\
 $1^{-}(2^{++})$ & $a_2 (1320)$
&$1308.72\pm22$&$0.62\%$&$1.91$ 
 \\
 $1^{+}(3^{--})$& $\rho_3 (1690)$
&$1679.12\pm31$&$0.57\%$&$3.03$
 \\
 $1^{-}(4^{++})$ & $a_4 (1970)$
&$1986.39\pm37$&$0.98\%$&$3.83$
 \\
 $1^{+}(5^{--})$& $\rho_5 (2350)$
&$2259.66\pm43$&$3.02\%$&$4.44$
 \\
 $1^{-}(6^{++})$& $a_6 (2450)$
&$2511.59\pm47$&$2.51\%$&$4.95$
\\ \hline
 $0^{-}(1^{--})$ & $\omega (782)$
&$782.66\pm10.17$&$0\%$&$0$
 \\
 $0^{+}(2^{++})$&$f_2 (1270)$
&$1303.83\pm22$&$2.22\%$&$1.82$
 \\
 $0^{-}(3^{--})$&$\omega_3 (1670)$
&$1668.32\pm30$&$0.08\%$&$2.88$
 \\
 $0^{+}(4^{++})$& $f_4 (2050)$
&$1972.01\pm37$&$2.28\%$&$3.63$
 \\
 $0^{-}(5^{--})$& $\omega_5 (2250)$
&$2242.93\pm42$&$0.31\%$&$4.22$
 \\
 $0^{+}(6^{++})$& $f_6 (2510)$
&$2493.25\pm46$&$1.15\%$&$4.69$
\\ \hline
 $1^{+}(1^{+-})$& $b_1 (1235)$
&$1229.5\pm16$&$0\%$&$0$
 \\
 $1^{-}(2^{-+})$&$\pi_2 (1670)$
&$1638.56\pm27$&$1.92\%$&$0.38$
 \\
 $1^{+}(3^{+-})$ & $b_3 (2030)$
&$1988.45\pm37$&$2.14\%$&$0.60$
 \\
 $1^{-}(4^{-+})$& $\pi_4 (2250)$
&$2312.88\pm44$&$2.79\%$&$0.76$
\\ \hline
 $0^{-}(1^{+-})$ & $h_1 (1170)$
&$1166\pm15$ &$0\%$&$0$
 \\
 $0^{+}(2^{-+})$ & $\eta_2 (1645)$ 
&$1629.93\pm25$ &$0.80\%$  &$0.66$ 
 \\
 $0^{-}(3^{+-})$ & $h_3 (2025)$
&$2003.12\pm33$ &$1.08\%$ &$1.04$ 
 \\
 $0^{+}(4^{-+})$& $\eta_4 (2330)$
&$2339.01\pm39$ &$0.47\%$  &$1.32$ 
 \\
 \hline 
 \hline 
\end{tabular}
\caption{\label{HWlogchi2} Masses in MeV for mesons  classified under even or odd $PC$  with isospins  $I=0,1$ and spins $S\ge1$, from the AHW Log model, determined from Eq. \eqref{masstower2}, and the percentage deviations $\delta_{\rm Log}= \delta_{i}\times 100\%$, with  $\delta_{i}$ given by Eq. \eqref{percentualdev}, of the masses from this model compared with PDG data, using  minimization with respect to the parameter   $\chi^2$, Eq. \eqref{chiquadrado}. For each family the mass of the state with $J=S=1$  is taken as an input from PDG. In the last column we show the anomalous dimension $\Delta_{\rm Log}$ for each state determined by Eq. \eqref{anomdim}.}
\end{table*}

We present here the more standard $\chi^2$ minimization as a complement to the $Q$  minimization discussed in section \ref{AHW}, and also used in section \ref{LHW} to obtain the best fit of PDG data \cite{PDG:2022pth}. As a second method to obtain the best fit of PDG data from the AHW models, we use the $\chi^2$ definition given by
\begin{equation}\label{chiquadrado}
   \chi^2= \sum_i \frac{|m_i-M_i|^2}{M_i},
\end{equation}
where $m_i$ is the mass obtained by our model and $M_i $ is the mass from PDG for a given state. 
So, according to this method, the best fit is such that $\chi^2$ given by (\ref{chiquadrado}) is minimum.

\begin{table}[h]
\begin{tabular}{|c|c|c|c|c|}
\hline\hline
$I^{G}(J^{PC})$  & Meson & $a_{\rm Log}$  & AHW (Log) Regge Trajectory (GeV) &$R^2$
\\ \hline
 $1^{+}(1^{--})$  & $\rho (770)$ &    
$2.76\pm0.12$ & 
 $J=(0.88\pm0.01)M^2 +0.50\pm0.03$&0.9997
\\
\hline
 $0^{-}(1^{--})$ & $\omega (782)$ &  
$2.62\pm0.11$ &
 $J=(0.89\pm0.01)M^2 +0.48\pm0.03$&0.9997 
\\ \hline
 $1^{+}(1^{+-})$& $b_1 (1235)$ & 
$0.55\pm0.09$ &
 $J=(0.78\pm0.02)M^2-0.14\pm0.08$ &0.9985 
\\ \hline
 $0^{-}(1^{+-})$& $h_1 (1170)$ & 
$0.95\pm0.07$ & 
 $ J=(0.73\pm0.01)M^2+0.04\pm0.05$ &0.9993 
\\
 \hline 
 \hline 
\end{tabular}
\caption{\label{Tlogchi2}The Regge trajectory in GeV units for each light unflavored meson family (same $PC$ parity and same isospin $I$) from the AHW model, Eqs. \eqref{anomdim} and \eqref{masstower2}, using  minimization with respect to the  parameter  $\chi^2$, Eq. \eqref{chiquadrado}. In the last column, we also show the value of $R^2$, defined in Eq. \eqref{Rquadrado}, for each trajectory.}
\end{table}

The results for the AHW model with logarithm anomalous dimensions, Section \ref{AHW}, now minimized by the $\chi^2$ method are presented in Tables \ref{HWlogchi2} and \ref{Tlogchi2}, and can be compared with the results obtained from the minimization of $Q$ shown in Tables \ref{HWlog} and \ref{Tlog}. A comparison of these tables shows that the results of the two minimization methods are very close to each other for each family, and actually coincide for the $h_1(1170)$ family.

Analogously, the minimization of $\chi^2$ can also be applied to the linear AHW model, discussed in Section \ref{LHW}. The results in this case are displayed in Tables \ref{linearTchi2} and \ref{Tlinchi2}. A comparison of these results with the ones from the $Q$ minimization, presented in Tables \ref{linearT} and \ref{Tlin}, shows that both methods produce similar outputs,
an even exactly equal for the $\omega(782)$ family.

 \begin{table*}[ht!]
\begin{tabular}{|c|c|c|c|c|}
\hline 
\hline 
$I^{G}(J^{PC})$ & Meson &  Linear AHW ($\chi^2$)&$\delta_{\rm LAHW}$&$\Delta_{\rm LAHW}$
\\ 
\hline
 $1^{+}(1^{--})$ & $\rho (770)$ &  $775.26\pm10.08$&$0\%$&$0$
 \\
 $1^{-}(2^{++})$&$a_2 (1320)$
&$1295.28\pm20$&$1.64\%$&$1.84$
 \\
 $1^{+}(3^{--})$& $\rho_3 (1690)$
&$1676.08\pm29$&$0.75\%$&$3.01$
 \\
 $1^{-}(4^{++})$ & $a_4 (1970)$
&$1990.44\pm37$&$1.19\%$&$3.85$
 \\
 $1^{+}(5^{--})$& $\rho_5 (2350)$ 
&$2263.87\pm44$&$2.84\%$&$4.47$
 \\
 $1^{-}(6^{++})$& $a_6 (2450)$
&$2508.87\pm50$&$2.40\%$&$4.93$
\\ \hline
 $0^{-}(1^{--})$ & $\omega (782)$
 &  $782.66\pm10.17$&$0\%$&$0$
 \\
 $0^{+}(2^{++})$ & $f_2 (1270)$
&$1294.28\pm20$ &$1.47\%$  &$1.76$ 
 \\
 $0^{-}(3^{--})$& $\omega_3 (1670)$
&$1669.09\pm29$ &$0.13\%$  &$2.88$ 
 \\
 $0^{+}(4^{++})$& $f_4 (2050)$
&$1978.53\pm37$ &$1.96\%$  &$3.67$ 
 \\
 $0^{-}(5^{--})$& $\omega_5 (2250)$
&$2247.67\pm44$ &$0.10\%$  &$4.24$ 
 \\
 $0^{+}(6^{++})$& $f_6 (2510)$
&$2488.81\pm50$ &$0.97\%$  &$4.67$ 
\\ \hline
 $1^{+}(1^{+-})$& $b_1 (1235)$
&$1229.5\pm16$&$0\%$&$0$
 \\
 $1^{-}(2^{-+})$& $\pi_2 (1670)$
&$1665.95\pm26$&$0.28\%$&$0.48$ 
 \\
 $1^{+}(3^{+-})$& $b_3 (2030)$
&$1998.25\pm36$&$1.66\%$&$0.64$
 \\
 $1^{-}(4^{-+})$& $\pi_4 (2250)$
&$2277.85\pm45$&$1.24\%$&$0.64$
\\ \hline
 $0^{-}(1^{+-})$& $h_1 (1170)$
&$1166\pm15$&$0\%$&$0$
 \\
 $0^{+}(2^{-+})$&$\eta_2 (1645)$
&$1649.35\pm26$&$2.00\%$&$0.73$
 \\
 $0^{-}(3^{+-})$ & $h_3 (2025)$
&$2014.27\pm36$&$0.53\%$&$1.09$
 \\
 $0^{+}(4^{-+})$ & $\eta_4 (2330)$
&$2320.03\pm45$&$0.34\%$&$1.24$
 \\
 \hline 
 \hline 
\end{tabular}
\caption{\label{linearTchi2} Masses in MeV for mesons  classified under even or odd $PC$ with isospins $I=0,1$ and spins $S\ge1$, from the Linear AHW model, determined from Eq. \eqref{masstower3}, and the percentage deviations $\delta_{\rm LAHW}=\delta_{i} \times 100\%$, with $\delta_{i}$ from Eq. \eqref{percentualdev},  of the masses from this model compared with PDG data,  using  minimization with respect to the  parameter  $\chi^2$, Eq. \eqref{chiquadrado}. For each family the mass of the state with $J=S=1$  is taken as an input. In the last column we show the total  dimension $\Delta_{\rm LAHW}$ for each state determined by Eq. \eqref{lineardim}.}
\end{table*}

\begin{table}[h]
\footnotesize
\begin{tabular}{|c|c|c|c|c|c|c|c|c|c|}
\hline\hline
$I^{G}(J^{PC})$ & Meson & $a_{\rm LAHW}$ & $\epsilon$ & Linear AHW  Regge Trajectory (GeV) & $n$&$R^2$
\\ \hline
  $1^{+}(1^{--})$ & $\rho (770)$ &    
$6.85\pm0.16$ &   
$0.00\pm0.005$ & 
 $J=(0.88\pm0.01)M^2 +0.51\pm0.02$ 
 & 6&0.999799
 \\
   & &  &  &  
 $J=(0.8524\pm0.0002)M^2 +0.60\pm0.01$ 
 & 50 &0.999997
 \\
\hline
 $0^{-}(1^{--})$ & $\omega (782)$ &  
$6.67\pm 0.13$ &
$0.00\pm0.005$ & 
 $J=(0.89\pm0.01)M^2 +0.49\pm0.02$ 
 & 6&0.999845 
\\ 
  & & & & 
 $J=(0.8766\pm0.0003)M^2 +0.52\pm0.01$ 
 & 50&0.999995  
 \\
\hline
 $1^{+}(1^{+-})$& $b_1 (1235)$ & 
$3.18\pm 0.11$ &
 $0.05\pm0.005$& 
 $J=(0.82\pm0.01)M^2-0.25\pm0.03$ 
 & 4&0.999841
 \\
  & & & &  
 $J=(0.8595\pm0.0003)M^2-0.44\pm0.01$ 
 & 50 &0.999993
\\ \hline
 $0^{-}(1^{+-})$& $h_1 (1170)$ & 
$3.81\pm 0.12$ & 
 $0.04\pm0.005$ & 
 $ J=(0.75\pm0.03)M^2
 -0.02\pm0.01$ 
 & 4&0.999964
 \\
  & & & & 
 $ J=(0.7599\pm0.0003)M^2
 -0.04\pm0.01$ 
 & 50 &0.999994
 \\
 \hline 
 \hline 
\end{tabular}
\caption{\label{Tlinchi2}The Regge trajectories in GeV units for each light unflavored meson family (same $PC$ parity and same isospin $I$) from the Linear AHW model, Eq. \eqref{masstower3}, using the minimization with respect to the  parameter  $\chi^2$, Eq. \eqref{chiquadrado}. For each meson family we show two Regge trajectories corresponding to the linear regression of two sets of points with dimension $n$. In the last column, we also show the value of $R^2$, defined in Eq. \eqref{Rquadrado}, for each trajectory.}
\end{table}





\begin{thebibliography}{99}

\section*{References}



\bibitem{Aharony:1999ti}
O.~Aharony, S.~S.~Gubser, J.~M.~Maldacena, H.~Ooguri and Y.~Oz,
``Large N field theories, string theory and gravity,''
Phys. Rept. \textbf{323}, 183-386 (2000)
doi:10.1016/S0370-1573(99)00083-6
[arXiv:hep-th/9905111 [hep-th]].

\bibitem{Casalderrey-Solana:2011dxg}
J.~Casalderrey-Solana, H.~Liu, D.~Mateos, K.~Rajagopal and U.~A.~Wiedemann,
``Gauge/String Duality, Hot QCD and Heavy Ion Collisions,''
Cambridge University Press, 2014,
doi:10.1017/9781009403504
[arXiv:1101.0618 [hep-th]].

\bibitem{Zaanen:2015oix}
J.~Zaanen, Y.~W.~Sun, Y.~Liu and K.~Schalm,
``Holographic Duality in Condensed Matter Physics,''
Cambridge Univ. Press, 2015,
doi:10.1017/CBO9781139942492

\bibitem{Polchinski:2001tt}
J.~Polchinski and M.~J.~Strassler,
``Hard scattering and gauge / string duality,''
Phys. Rev. Lett. \textbf{88}, 031601 (2002)
doi:10.1103/PhysRevLett.88.031601
[arXiv:hep-th/0109174 [hep-th]].

\bibitem{Polchinski:2002jw}
J.~Polchinski and M.~J.~Strassler,
``Deep inelastic scattering and gauge / string duality,''
JHEP \textbf{05}, 012 (2003)
doi:10.1088/1126-6708/2003/05/012
[arXiv:hep-th/0209211 [hep-th]].


\bibitem{Boschi-Filho:2002xih}
H.~Boschi-Filho and N.~R.~F.~Braga,
``Gauge / string duality and scalar glueball mass ratios,''
JHEP \textbf{05}, 009 (2003)
doi:10.1088/1126-6708/2003/05/009
[arXiv:hep-th/0212207 [hep-th]].


\bibitem{Boschi-Filho:2002wdj}
H.~Boschi-Filho and N.~R.~F.~Braga,
``QCD / string holographic mapping and glueball mass spectrum,''
Eur. Phys. J. C \textbf{32}, 529-533 (2004)
doi:10.1140/epjc/s2003-01526-4
[arXiv:hep-th/0209080 [hep-th]].
 

\bibitem{deTeramond:2005su}
G.~F.~de Teramond and S.~J.~Brodsky,
``Hadronic spectrum of a holographic dual of QCD,''
Phys. Rev. Lett. \textbf{94}, 201601 (2005)
doi:10.1103/PhysRevLett.94.201601
[arXiv:hep-th/0501022 [hep-th]].
 
\bibitem{Erlich:2005qh}
J.~Erlich, E.~Katz, D.~T.~Son and M.~A.~Stephanov,
``QCD and a holographic model of hadrons,''
Phys. Rev. Lett. \textbf{95}, 261602 (2005)
doi:10.1103/PhysRevLett.95.261602
[arXiv:hep-ph/0501128 [hep-ph]].


\bibitem{Boschi-Filho:2005xct}
H.~Boschi-Filho, N.~R.~F.~Braga and H.~L.~Carrion, 
``Glueball Regge trajectories from gauge/string duality and the Pomeron,'' 
Phys. Rev. D \textbf{73}, 047901 (2006) 
doi:10.1103/PhysRevD.73.047901 
[arXiv:hep-th/0507063 [hep-th]].


\bibitem{Karch:2006pv}
A.~Karch, E.~Katz, D.~T.~Son and M.~A.~Stephanov,
``Linear confinement and AdS/QCD,''
Phys. Rev. D \textbf{74}, 015005 (2006)
doi:10.1103/PhysRevD.74.015005
[arXiv:hep-ph/0602229 [hep-ph]].

\bibitem{Colangelo:2008us}
P.~Colangelo, F.~De Fazio, F.~Giannuzzi, F.~Jugeau and S.~Nicotri,
``Light scalar mesons in the soft-wall model of AdS/QCD,'' 
Phys. Rev. D \textbf{78}, 055009 (2008)
doi:10.1103/PhysRevD.78.055009
[arXiv:0807.1054 [hep-ph]].

\bibitem{Boschi-Filho:2005nmp}
H.~Boschi-Filho, N.~R.~F.~Braga and C.~N.~Ferreira,
``Static strings in Randall-Sundrum scenarios and the quark anti-quark potential,''
Phys. Rev. D \textbf{73}, 106006 (2006)
[erratum: Phys. Rev. D \textbf{74}, 089903 (2006)]
doi:10.1103/PhysRevD.74.089903
[arXiv:hep-th/0512295 [hep-th]].

\bibitem{Boschi-Filho:2006hfm}
H.~Boschi-Filho, N.~R.~F.~Braga and C.~N.~Ferreira,
``Heavy quark potential at finite temperature from gauge/string duality,''
Phys. Rev. D \textbf{74}, 086001 (2006)
doi:10.1103/PhysRevD.74.086001
[arXiv:hep-th/0607038 [hep-th]].


\bibitem{Herzog:2006ra}
C.~P.~Herzog,
``A Holographic Prediction of the Deconfinement Temperature,''
Phys. Rev. Lett. \textbf{98}, 091601 (2007)
doi:10.1103/PhysRevLett.98.091601
[arXiv:hep-th/0608151 [hep-th]].

\bibitem{BallonBayona:2007vp}
C.~A.~Ballon Bayona, H.~Boschi-Filho, N.~R.~F.~Braga and L.~A.~Pando Zayas,
``On a Holographic Model for Confinement/Deconfinement,''
Phys. Rev. D \textbf{77}, 046002 (2008)
doi:10.1103/PhysRevD.77.046002
[arXiv:0705.1529 [hep-th]].

\bibitem{Grigoryan:2007my}
H.~R.~Grigoryan and A.~V.~Radyushkin,
``Structure of vector mesons in holographic model with linear confinement,''
Phys. Rev. D \textbf{76}, 095007 (2007)
doi:10.1103/PhysRevD.76.095007
[arXiv:0706.1543 [hep-ph]].

\bibitem{Kwee:2007dd}
H.~J.~Kwee and R.~F.~Lebed,
``Pion form-factors in holographic QCD,''
JHEP \textbf{01}, 027 (2008)
doi:10.1088/1126-6708/2008/01/027
[arXiv:0708.4054 [hep-ph]].

\bibitem{Grigoryan:2007wn}
H.~R.~Grigoryan and A.~V.~Radyushkin,
``Pion form-factor in chiral limit of hard-wall AdS/QCD model,''
Phys. Rev. D \textbf{76}, 115007 (2007)
doi:10.1103/PhysRevD.76.115007
[arXiv:0709.0500 [hep-ph]].

\bibitem{Jo:2009xr}
K.~Jo, B.~H.~Lee, C.~Park and S.~J.~Sin,
``Holographic QCD in medium: A Bottom up approach,''
JHEP \textbf{06}, 022 (2010)
doi:10.1007/JHEP06(2010)022
[arXiv:0909.3914 [hep-ph]].

\bibitem{Costa:2013uia}
M.~S.~Costa, M.~Djuri\'c and N.~Evans,
``Vector meson production at low x from gauge/gravity duality,''
JHEP \textbf{09}, 084 (2013)
doi:10.1007/JHEP09(2013)084
[arXiv:1307.0009 [hep-ph]].

\bibitem{Craps:2013iaa}
B.~Craps, E.~Kiritsis, C.~Rosen, A.~Taliotis, J.~Vanhoof and H.~b.~Zhang,
``Gravitational collapse and thermalization in the hard wall model,''
JHEP \textbf{02}, 120 (2014)
doi:10.1007/JHEP02(2014)120
[arXiv:1311.7560 [hep-th]].

\bibitem{Mamo:2015dea}
K.~A.~Mamo,
``Inverse magnetic catalysis in holographic models of QCD,''
JHEP \textbf{05}, 121 (2015)
doi:10.1007/JHEP05(2015)121
[arXiv:1501.03262 [hep-th]].

\bibitem{Ballon-Bayona:2017bwk}
A.~Ballon-Bayona, G.~Krein and C.~Miller,
``Strong couplings and form factors of charmed mesons in holographic QCD,''
Phys. Rev. D \textbf{96}, no.1, 014017 (2017)
doi:10.1103/PhysRevD.96.014017
[arXiv:1702.08417 [hep-ph]].


\bibitem{Gubser:2002tv}
S.~S.~Gubser, I.~R.~Klebanov and A.~M.~Polyakov,
``A Semiclassical limit of the gauge / string correspondence,''
Nucl. Phys. B \textbf{636}, 99-114 (2002)
doi:10.1016/S0550-3213(02)00373-5
[arXiv:hep-th/0204051 [hep-th]].

\bibitem{Costa-Silva:2023vuu}
R.~A.~Costa-Silva and H.~Boschi-Filho,
``Anomalous and linear holographic hard wall models for glueballs and the pomeron,''
Phys. Rev. D \textbf{109}, no.8, 086019 (2024)
doi:10.1103/PhysRevD.109.086019
[arXiv:2306.04728 [hep-ph]].


\bibitem{PDG:2022pth}
R.~L.~Workman \textit{et al.} [Particle Data Group],
``Review of Particle Physics,''
PTEP \textbf{2022}, 083C01 (2022)
doi:10.1093/ptep/ptac097


\bibitem{Chen:2021kfw}
J.~K.~Chen,
``Structure of the meson Regge trajectories,''
Eur. Phys. J. A \textbf{57}, no.7, 238 (2021)
doi:10.1140/epja/s10050-021-00502-y
[arXiv:2102.07993 [hep-ph]].

\bibitem{Spector:1967xsn}
R.~M.~Spector,
``Example of a Regge daughter trajectory for $I$ = 1, $Y$ = 0 boson resonances,''
Phys. Lett. B \textbf{25}, 551-553 (1967)
doi:10.1016/0370-2693(67)90144-X

\bibitem{Kanki:1975fp}
T.~Kanki,
``Chew-Frautschi Plot and String Structure of Baryons,''
Lett. Nuovo Cim. \textbf{15}, 463 (1976)
doi:10.1007/BF02725148

\bibitem{Operador}
It is also possible to alternatively write the mesons with spin $S$ as  \cite{deTeramond:2005su}
 %
 \begin{equation}\label{mesonOpalt}
     {\cal O}^{\mu \,\mu_1 ... \mu_S}_{2+S}= 
     \;\bar\psi \gamma^\mu 
     D_{\{\mu_1\dots}D_{\mu_S\}}\psi\;;
     \qquad(S=1, 2, 3, ...)\,,
 \end{equation}
 %
 such that 
%
\begin{equation}\label{canonicaldim2+2}
     \Delta=2+S. 
 \end{equation}
 %
 However, in this case, the results for the AHW and LAHW are not as good as the ones discussed in Sections \ref{AHW} and \ref{LHW},   with $\Delta=3+S$.

\end{thebibliography}
\end{document}